\documentclass[usenatbib]{mn2e}
\usepackage{psfig,rotating}
\usepackage{graphicx,color}
\usepackage{dcolumn}
\usepackage{natbib}
\DeclareMathAlphabet{\mathsc}{OT1}{cmr}{m}{sc}
\def\testbx{bx}%
\DeclareRobustCommand{\ion}[2]{%
\relax\ifmmode
\ifx\testbx\f@series
{\mathbf{#1\,\mathsc{#2}}}\else
{\mathrm{#1\,\mathsc{#2}}}\fi
\else\textup{#1\,{\mdseries\textsc{#2}}}%
\fi}

\def\h2{\ensuremath{\rm H_2}}
\def\lya{\ensuremath{{\rm Ly}\alpha}}

\def\kms{km\,s$^{-1}$}

\newcommand{\be}{\begin{equation}}
\newcommand{\en}{\end{equation}}

\def\zabs{$z_{\rm abs}$}
\def\zem{$z_{\rm em}$~}
\def\lya{Ly$\alpha$ }

\def\hi{H~{\sc i}~}

\def\kms{km~s$^{-1}$}
\title[21-cm absorption in DLAs]
{Search for cold gas in $z>2$ damped \lya systems: 21-cm and H$_2$ absorption}
\author[Srianand et al.] {R. Srianand$^{1}$\thanks{E-mail:anand@iucaa.ernet.in}, N. Gupta$^{2}$, P. Petitjean$^3$, P. Noterdaeme$^3$, C. Ledoux$^4$, 
\newauthor C. J. Salter$^5$ and D. J. Saikia$^6$\\
$^{1}$ IUCAA, Ganeshkhind, Pune 411007, India \\
$^{2}$ Netherlands Institute for Radio Astronomy  (ASTRON), Postbus 2, 7990 AA, Dwingeloo, The Netherlands\\
$^{3}$ Universit\'e Paris 6, UMR 7095, Institut d'Astrophysique de Paris-CNRS, 98bis Boulevard Arago, 75014 Paris, France \\
$^4$ European Southern Observatory, Alonso de C\'ordova 3107, Casilla
19001, Vitacura, Santiago 19, Chile\\
$^5$ Arecibo Observatory, NAIC, HC3 Box 53995, Arecibo, Puerto Rico, PR 00612, USA\\
$^{6}$ NCRA-TIFR, Ganeshkhind, Pune 411007, India }

\begin{document}

\date{Accepted. Received; in original form }

\pagerange{\pageref{firstpage}--\pageref{lastpage}} \pubyear{2011}

\maketitle

\label{firstpage}

\begin{abstract}
We present the results of a systematic Green Bank Telescope (GBT) and 
Giant Metrewave Radio Telescope (GMRT) survey for 21-cm absorption in a 
sample of 10 damped Lyman-$\alpha$ (DLA) systems at $2\le$ \zabs$\le 3.4$.  
Analysis of L-band Very Long Baseline Array (VLBA) images of the
background QSOs are also presented.  We detect
21-cm absorption in only one DLA (at \zabs\ = 3.1745 towards J1337+3152).
Thus the detection rate of 21-cm absorption is  $\sim 10\%$ when no 
limit on the integrated optical depth ($\int \tau(v) dv$) 
is imposed and $\sim13$\% for a 3$\sigma$ limit of 0.4~\kms. 
Combining our data with the data from the literature (a sample
of 28 DLAs) and assuming the measured core fraction at milliarcsecond 
scale to represent the gas covering factor, we find that the H~{\sc i} gas 
in DLAs at $z\ge 2$ is predominantly constituted by warm neutral medium.
The detection rate of 21-cm absorption seems to be higher for systems with 
higher $N$(H~{\sc i}) or metallicity. However,
no clear correlation is found between the integrated 21-cm optical depth (or 
the spin-temperature, $T_{\rm S}$) and either  $N$(H~{\sc i}), metallicity 
or velocity spread of the low ionization species.
There are 13 DLAs in our sample for which high resolution optical
spectra covering the expected wavelength range of \h2 absorption are
available.
We report the detection of \h2 molecules in the \zabs\ ~=~3.3871 21-cm absorber 
towards J0203+1134 (PKS 0201+113).
In 8 cases, neither \h2 (with molecular fraction $f$($H_2$)$ \le 10^{-6}$)
nor 21-cm absorption (with $T_{\rm S}/f_{\rm c} \ge$~700~K) are detected. 
The lack of 21-cm and H$_2$ absorption in these systems can be 
explained if most of the H~{\sc i} in these DLAs originate from low density
high temperature gas.
{In one case we have a DLA with 21-cm absorption not showing \h2 absorption.}
In two cases, both species are detected but do not 
originate from the same velocity component. In the remaining 2 cases
21-cm absorption is not detected despite the presence of \h2 with evidence for the presence of cold gas. 
All this is consistent with the idea that the \h2 components seen in DLAs are compact (with sizes of $\le$~15~pc) 
and contain only a small fraction (i.e typically $\le 10\%$) of the total $N$(H~{\sc i}) measured in the DLAs. 
This implies that the molecular fractions $f$(\h2) reported from the \h2 surveys should be considered as conservative 
lower limits for the \h2 components.

\end{abstract}
%

\section{Introduction}

The Galactic interstellar medium (ISM) has a multiphase structure 
with neutral hydrogen being distributed between the 
cold neutral (CNM), warm neutral (WNM) and warm 
ionized (WIM) media. A large fraction of the gas is also found in 
diffuse, translucent and dense molecular clouds. Newly formed stars
are associated with these dense molecular clouds and 
strongly influence the physical state of  the rest of the gas in different
forms through radiative and mechanical inputs. 
The physical conditions in the multiphase ISM depend on the UV background 
radiation field, metallicities, dust content and the
density of cosmic rays 
\citep[see  Figs.~5, 6 and 7 in][]{Wolfire95}.
In addition, 
the filling factor of the different phases depends sensitively
on the supernova rate \citep{deavillez2004}.
Therefore, detecting and studying the multiphase ISM
in external galaxies has great importance for our understanding of 
galaxy evolution.

Damped Lyman-$\alpha$ systems (DLAs) are the highest H~{\sc i} column
density  absorbers seen in QSO spectra, with 
$N$(H~{\sc i})$\ge 2\times 10^{20}$~cm$^{-2}$.
These absorbers trace the bulk of the neutral hydrogen at $2\le z\le 3$ 
\citep [][]{Prochaska05,Noterdaeme09dla} and have long been identified as 
revealing the interstellar medium of the high-redshift precursors
of present day galaxies \citep[for a review see,][]{wolfe05}.

The typical dust-to-gas ratio of DLAs, is less than one tenth of that observed 
in the local ISM, and only a small fraction ($<∼10$\%) of DLAs show detectable
amounts of molecular hydrogen \citep{Petitjean00,Ledoux03,Noterdaeme08} with
the detection rate being correlated to the dust content of the gas 
\citep{Petitjean06}. 
The estimated temperature and molecular fraction in these 
systems are consistent with them 
originating from the CNM \citep{Srianand05}.
It has been shown recently that strong C~{\sc i} absorbers detected in 
low-resolution Sloan Digital Sky Survey (SDSS) 
spectra are good candidates for H$_2$ bearing systems.
Indeed these absorbers
have yielded the first detections of CO molecules in high-$z$ DLAs
\citep{Srianand08,Noterdaeme09co,Noterdaeme10co,Noterdaeme11}.
The properties of these absorbers are similar to those of translucent 
molecular clouds. 
The fact that no DLA is found to be associated with a dense molecular 
cloud, a fundamental ingredient 
of star-formation, is most certainly related to 
the large extinction that these clouds are expected to produce and/or
the small size of such regions \citep{Zwaan06} making 
detections difficult.

Thus, most DLAs detected in optical spectroscopic surveys seem to 
probe the diffuse H~{\sc i} 
gas \citep{Petitjean00}. However, 
about 50\% of the DLAs show detectable C~{\sc ii}$^*$ absorption
\citep{Wolfe08}, and \citet{Wolfe03b} argued that a 
considerable fraction of the
C~{\sc ii}$^*$ absorption in DLAs originates from CNM 
gas \citep[see however][]{Srianand05}. 
{ Detection of 21-cm absorption is the best way to estimate the CNM
fraction of DLAs as it is sensitive to both $N$(H~{\sc i}) and
thermal state of the gas \citep{Kulkarni88}.}
%
%
\begin{figure*}
\centerline{
\vbox{
\hbox{
\includegraphics[trim= 25.0mm 50.0mm 90.0 40.0mm, clip, scale=0.33,angle=90.0]{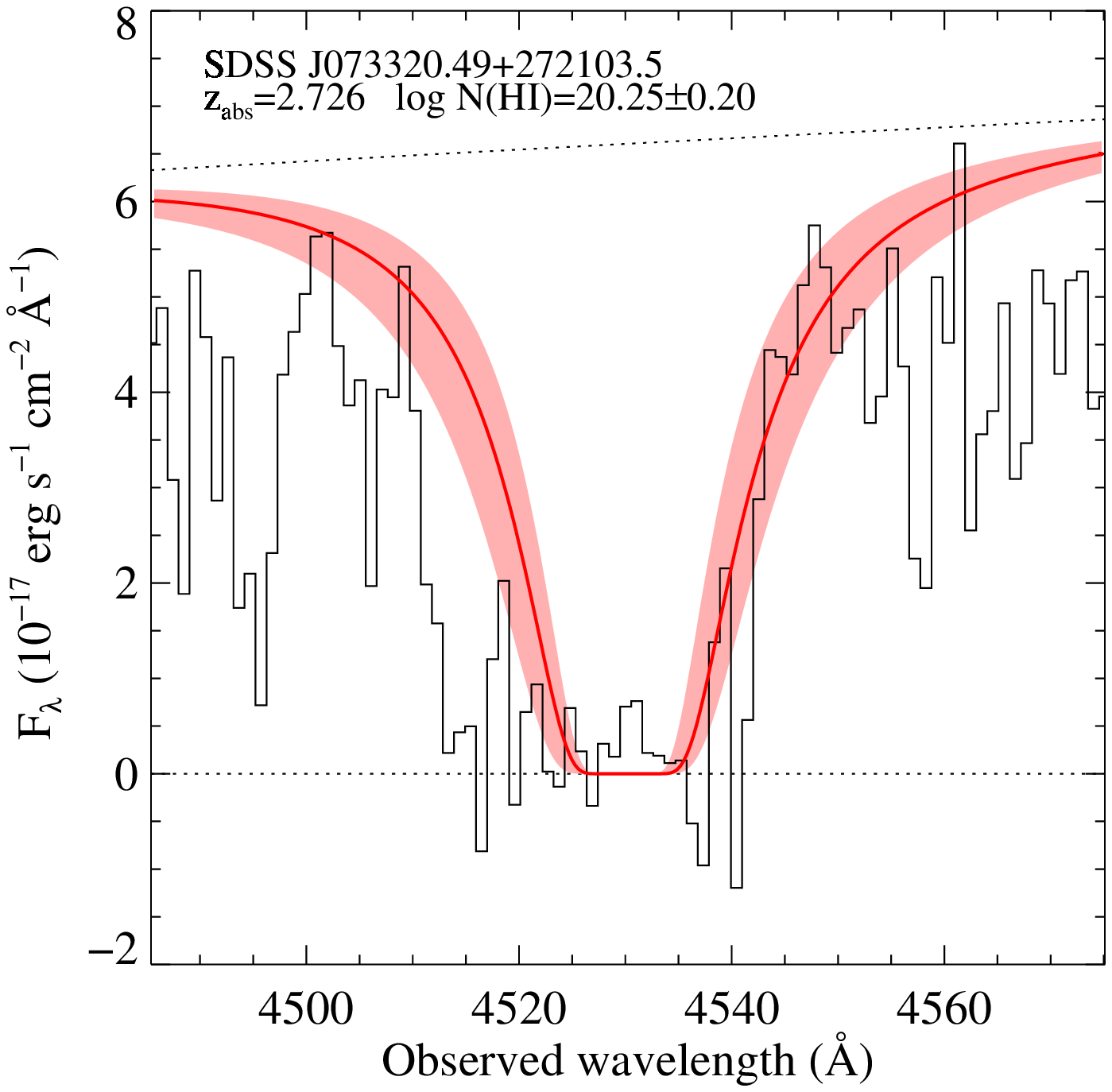}
\includegraphics[trim= 25.0mm 50.0mm 90.0 50.0mm, clip, scale=0.33,angle=90.0]{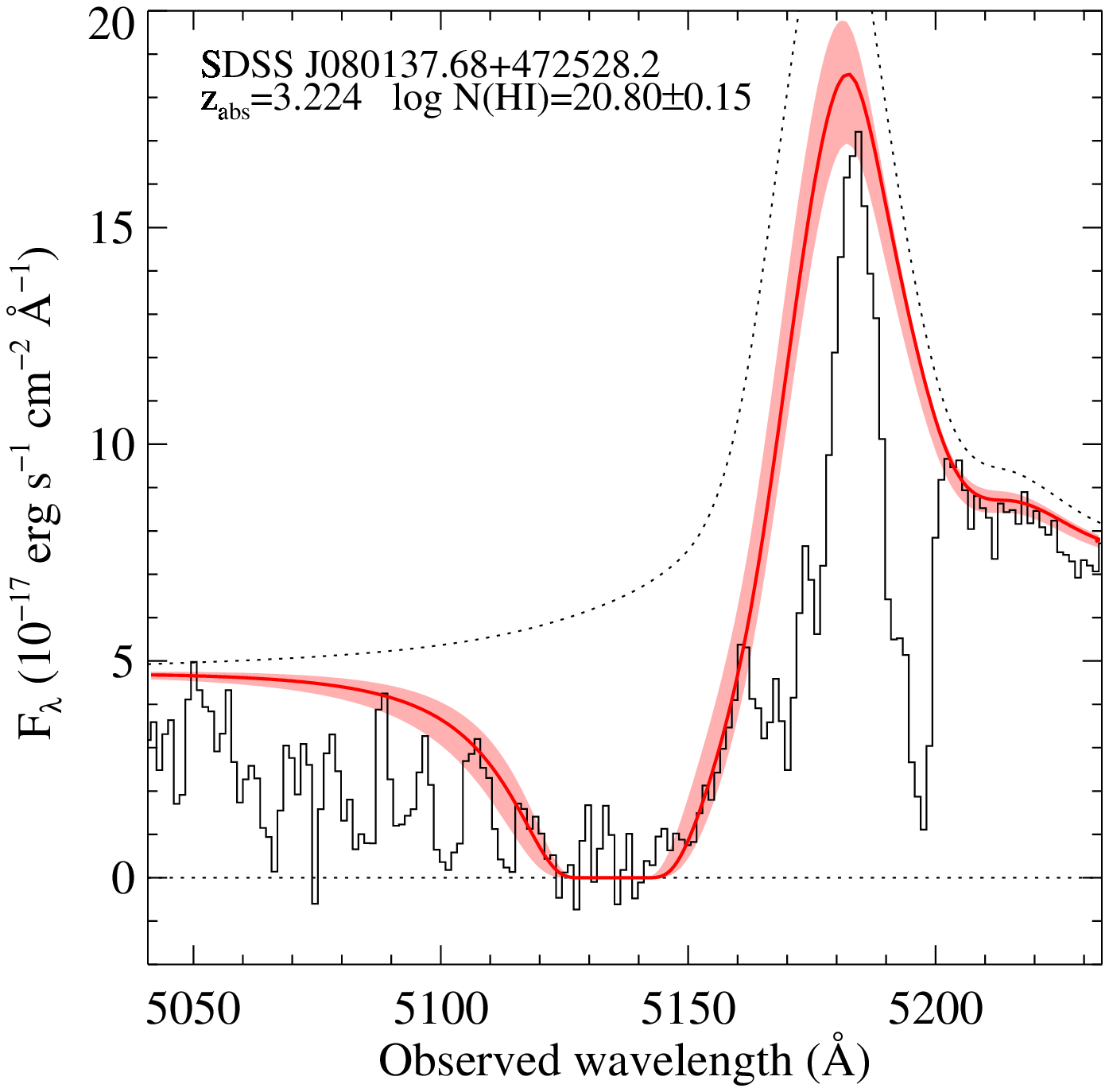}
\includegraphics[trim= 25.0mm 50.0mm 90.0 50.0mm, clip, scale=0.33,angle=90.0]{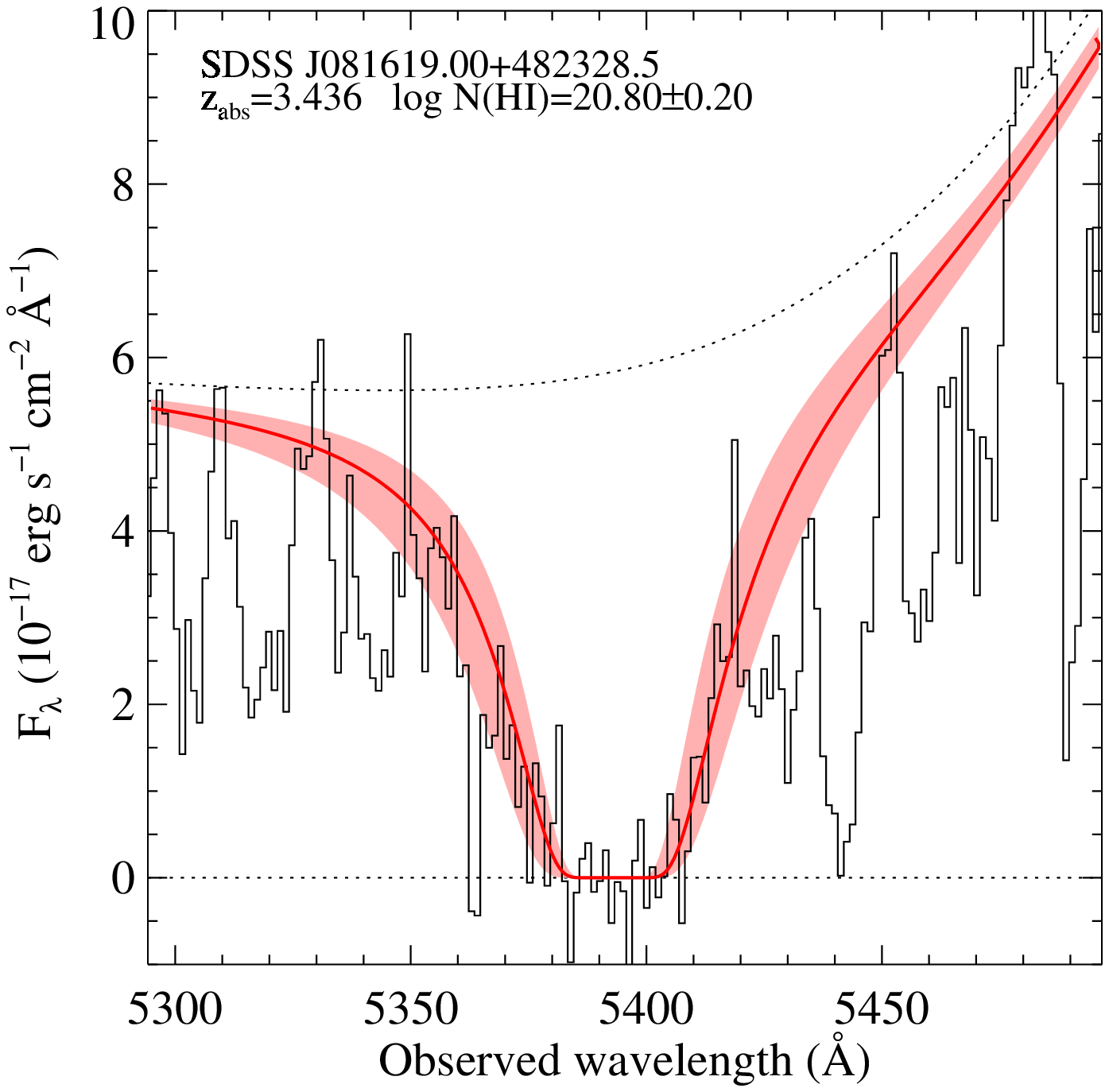}
}
\hbox{
\includegraphics[trim= 25.0mm 50.0mm 70.0 40.0mm, clip, scale=0.33,angle=90.0]{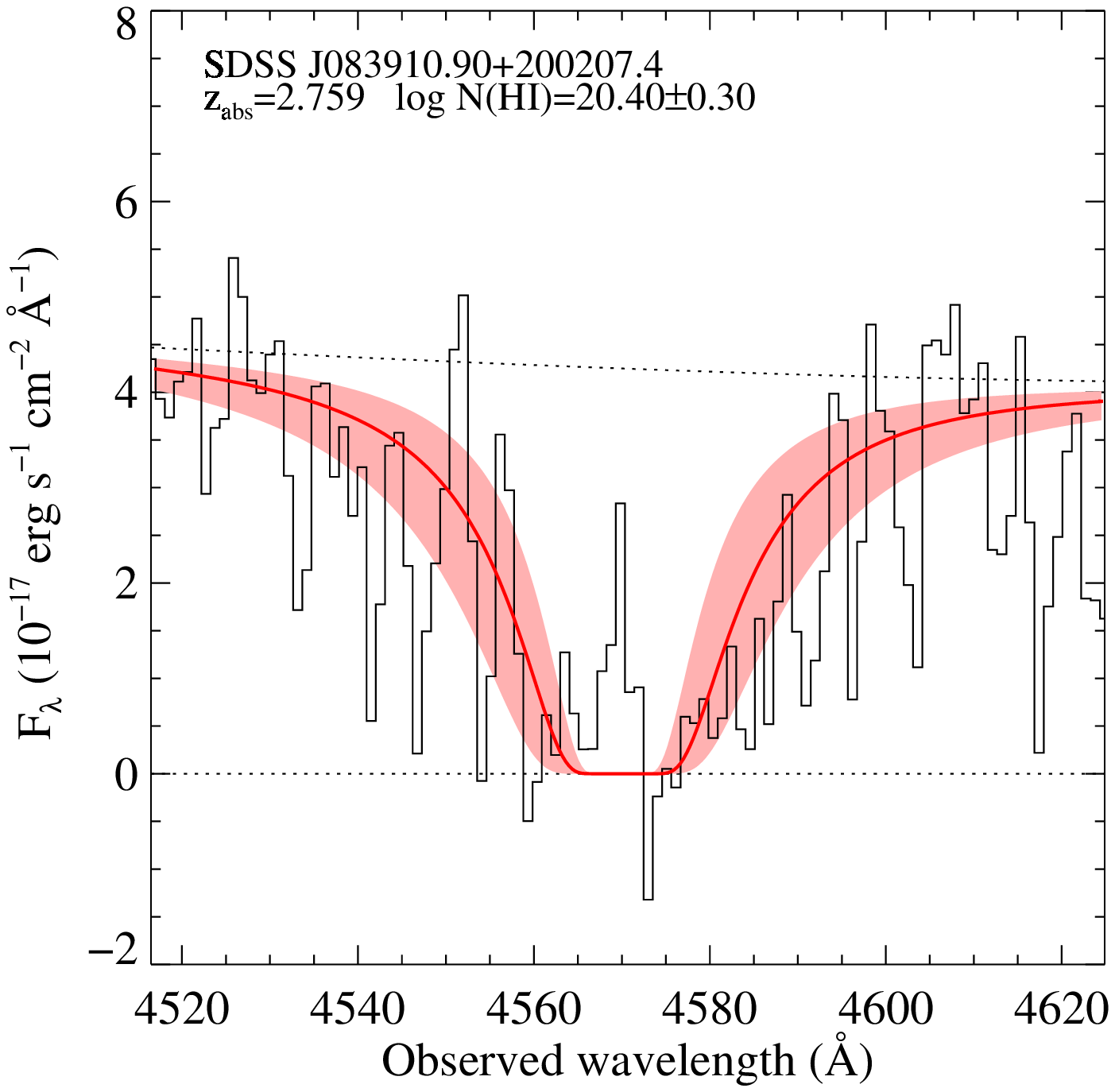}
\includegraphics[trim= 25.0mm 50.0mm 70.0 50.0mm, clip, scale=0.33,angle=90.0]{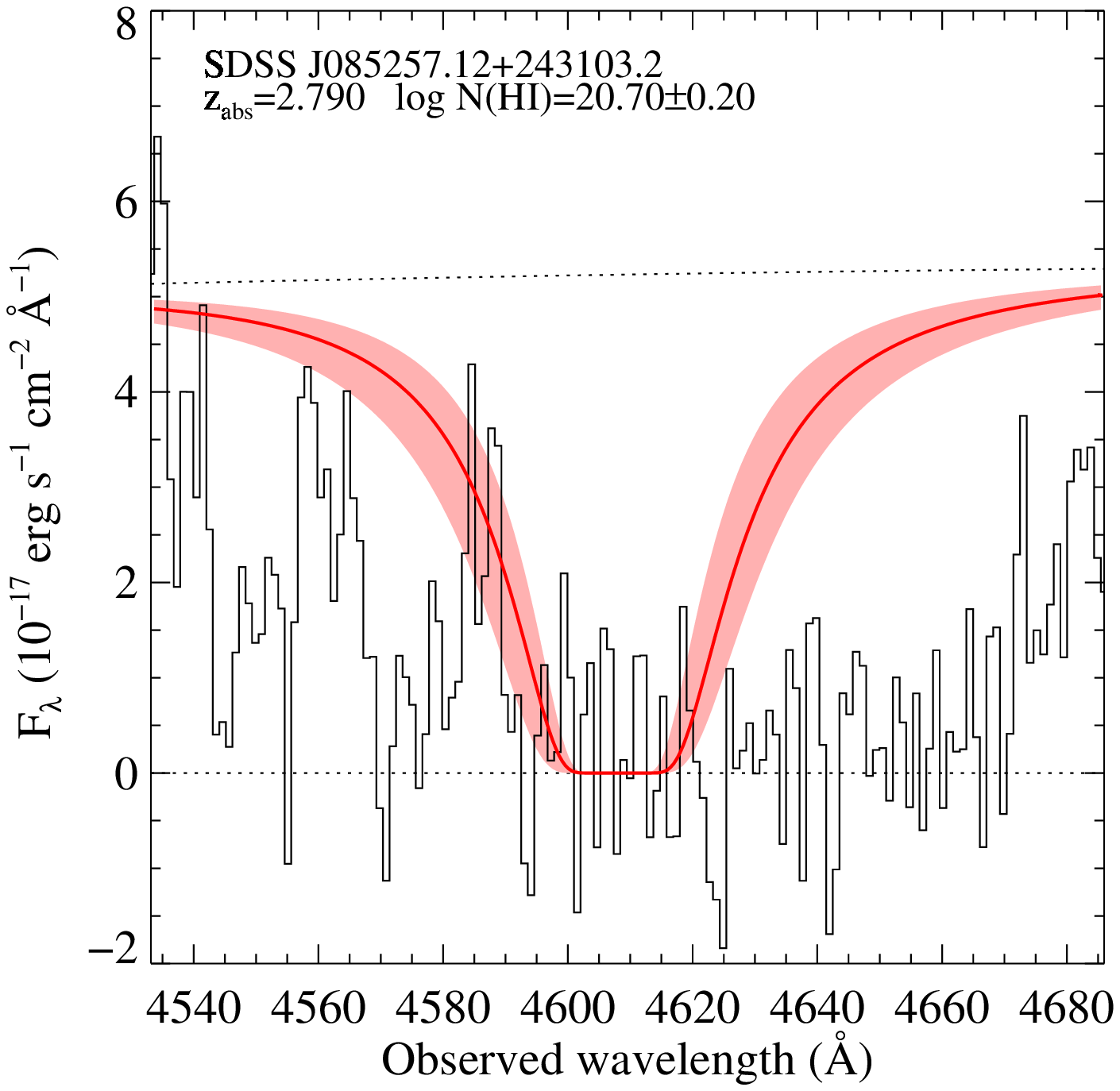}
\includegraphics[trim= 25.0mm 50.0mm 70.0 50.0mm, clip, scale=0.33,angle=90.0]{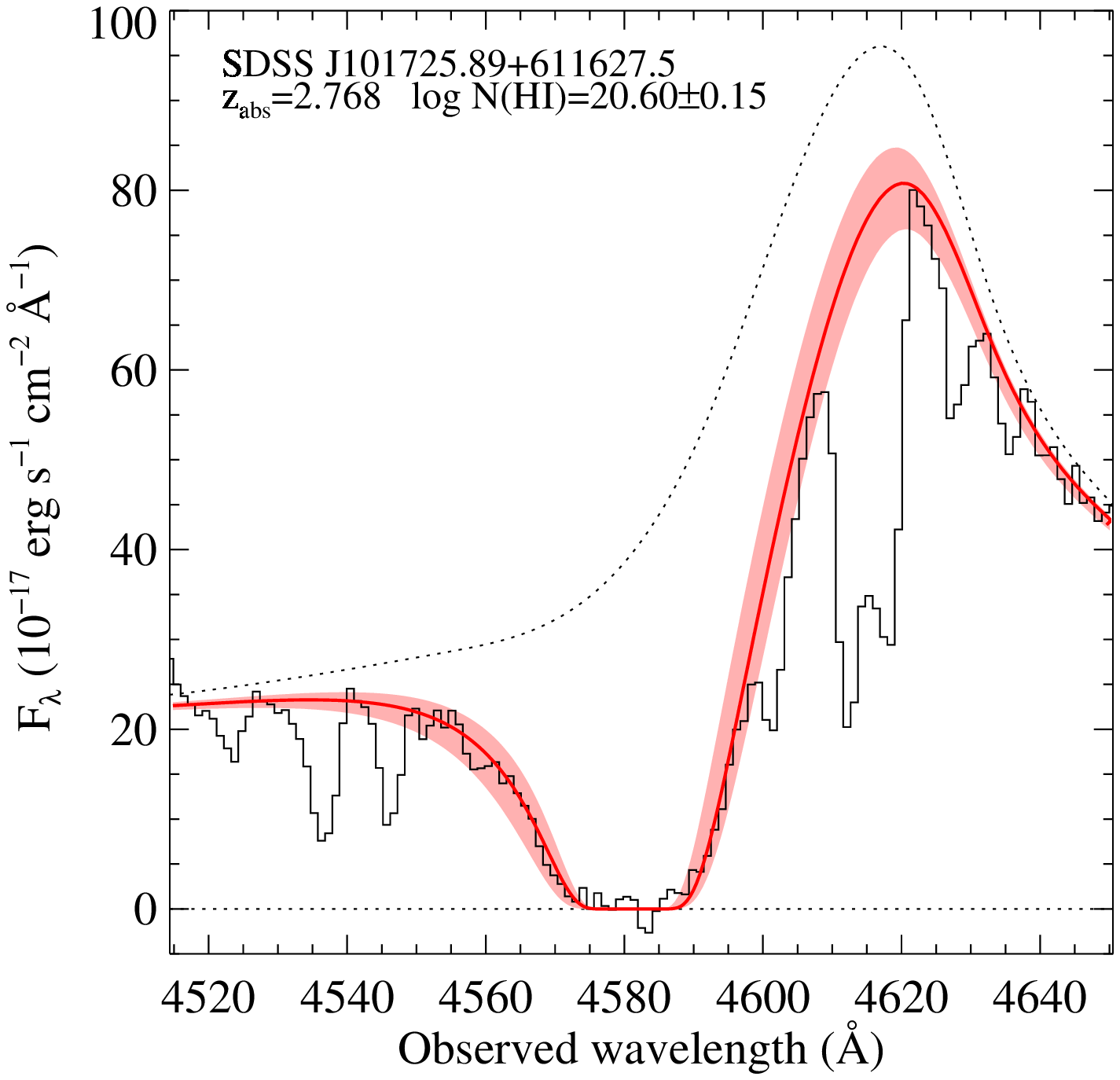}
}
\hbox{
\includegraphics[trim= 25.0mm 50.0mm 70.0 40.0mm, clip, scale=0.33,angle=90.0]{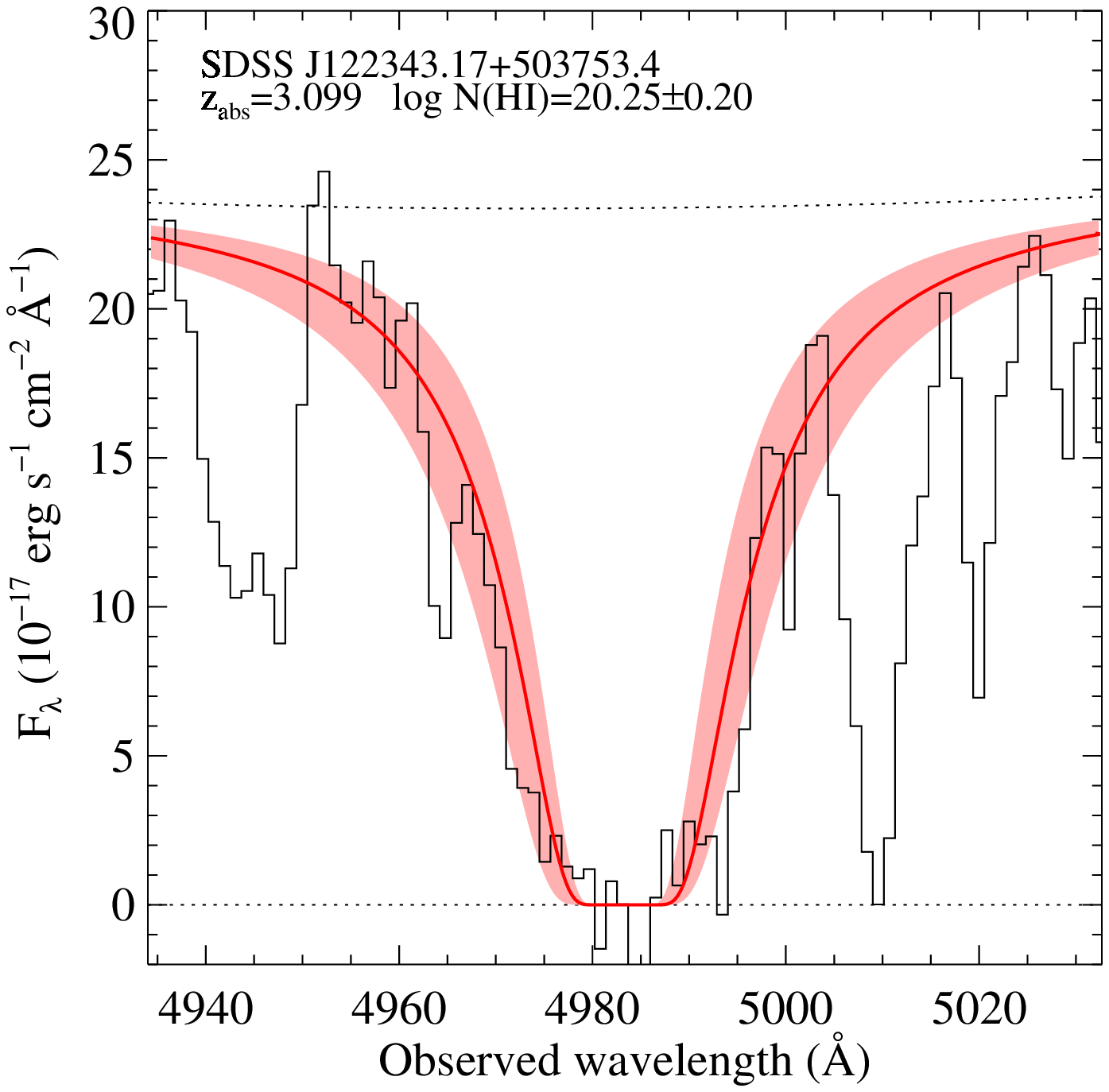}
\includegraphics[trim= 25.0mm 50.0mm 70.0 50.0mm, clip, scale=0.33,angle=90.0]{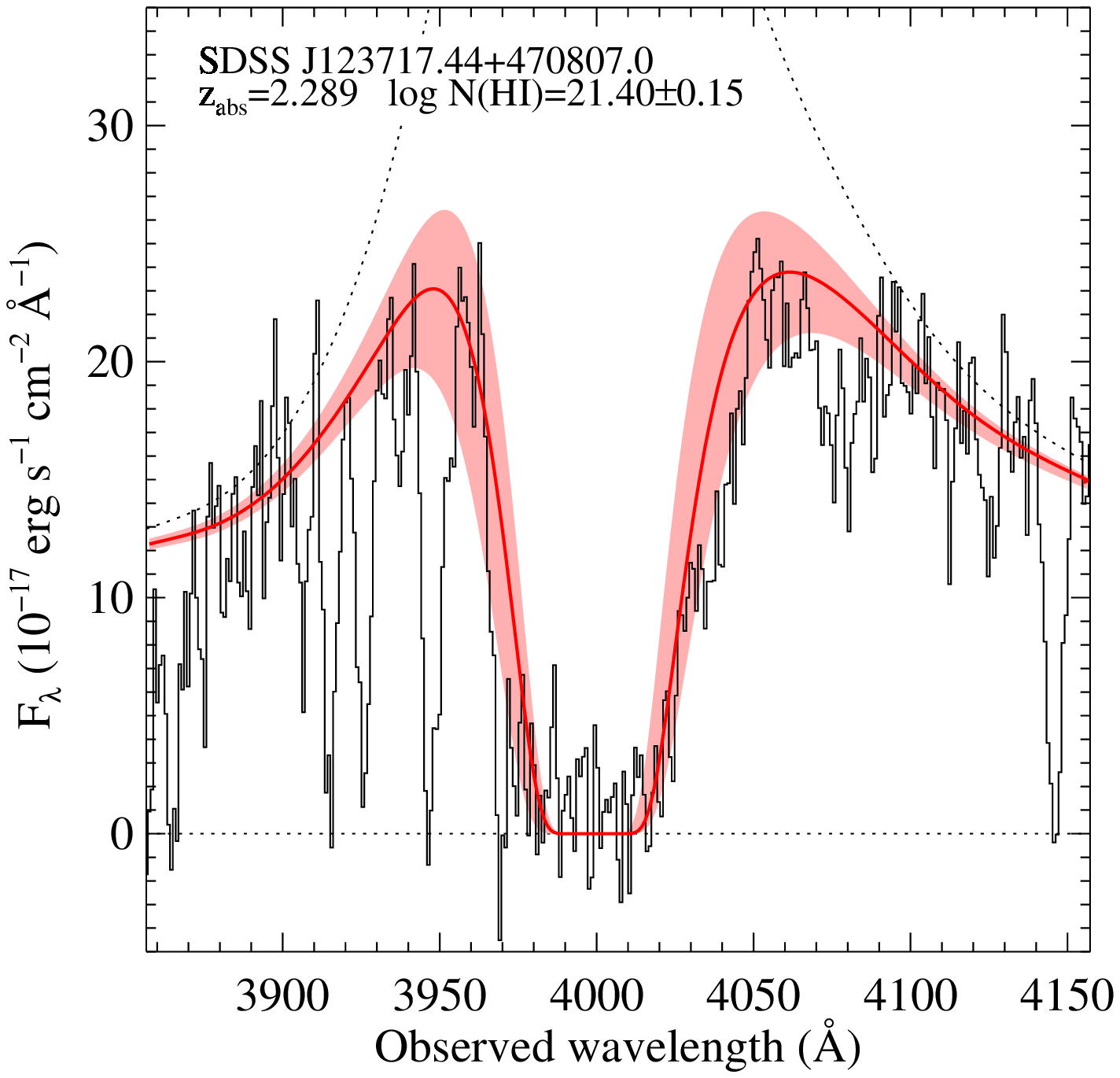}
\includegraphics[trim= 25.0mm 50.0mm 70.0 50.0mm, clip, scale=0.33,angle=90.0]{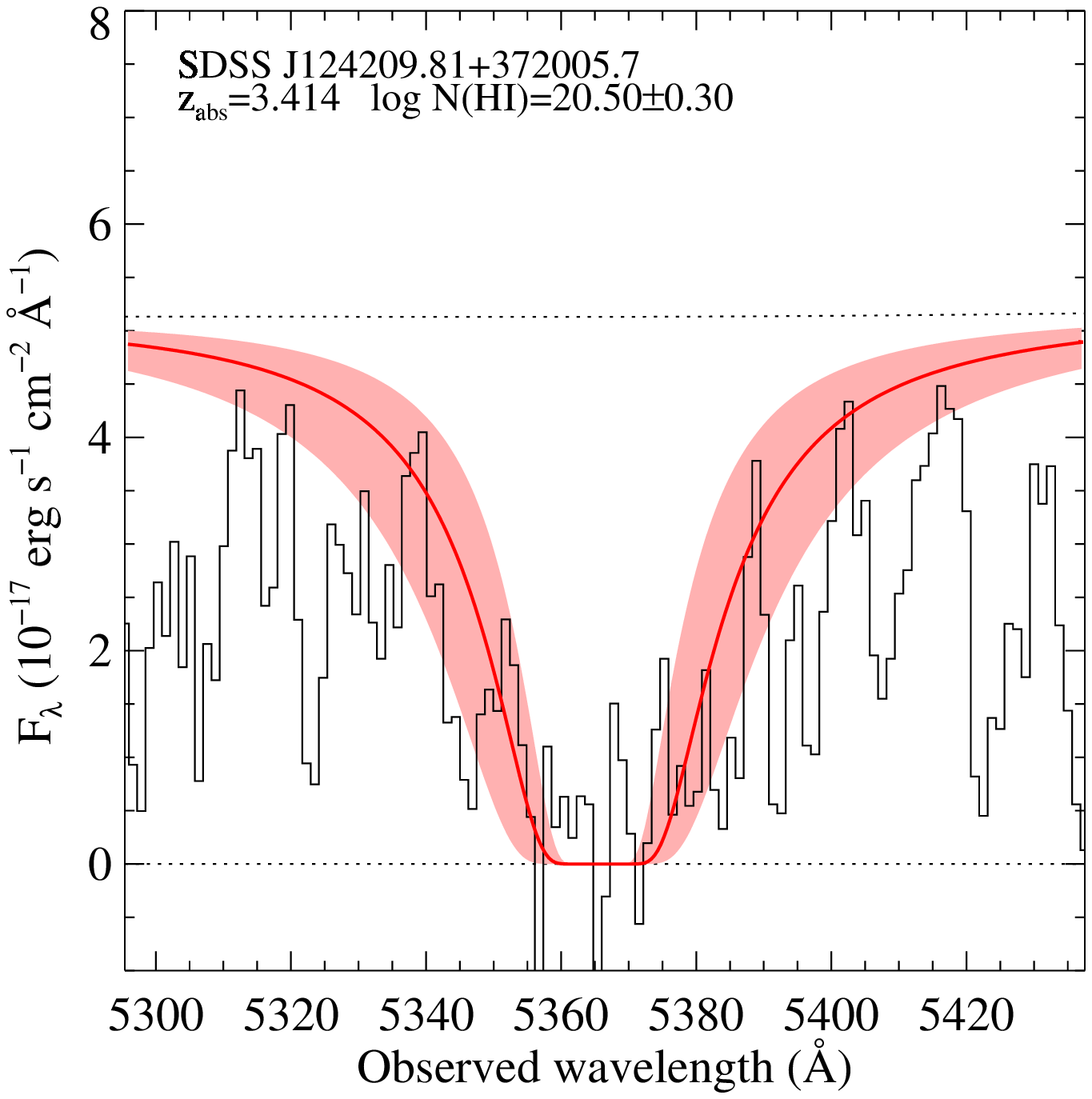}
}
\hbox{
\includegraphics[trim= 25.0mm 50.0mm 70.0 40.0mm, clip, scale=0.33,angle=90.0]{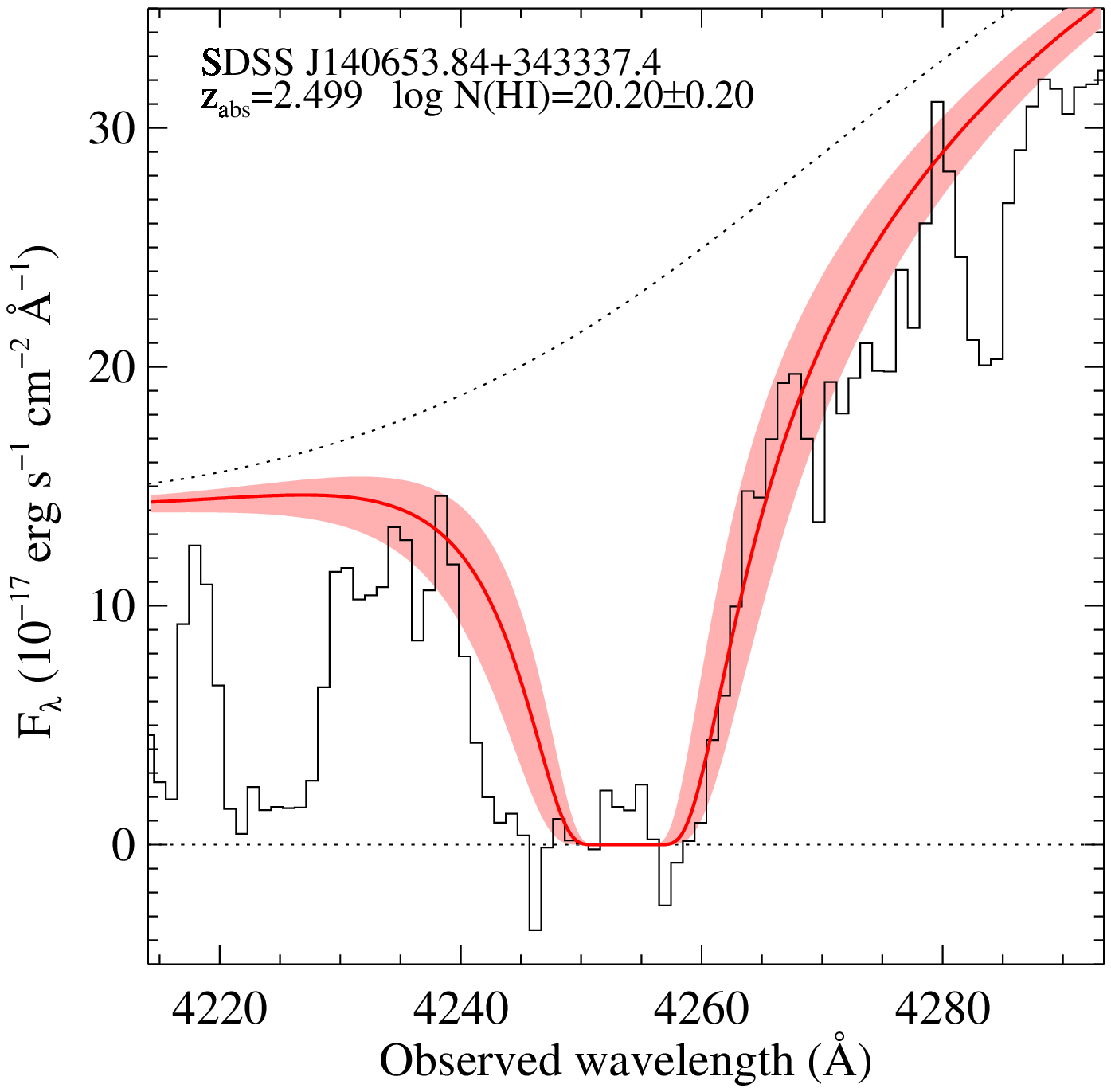}
\includegraphics[trim= 25.0mm 50.0mm 90.0 50.0mm, clip, scale=0.33,angle=90.0]{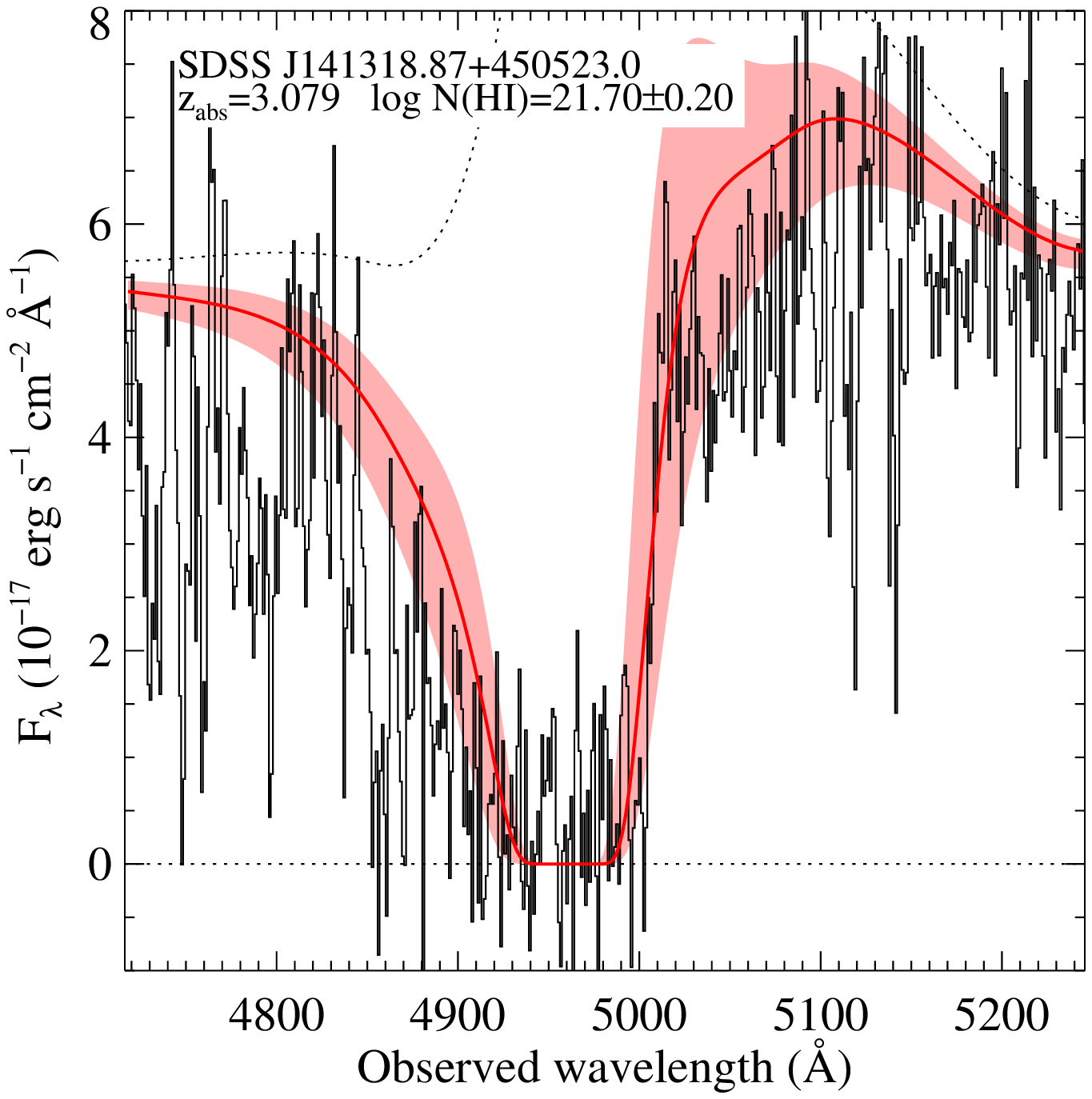}
\includegraphics[trim= 25.0mm 50.0mm 90.0 50.0mm, clip, scale=0.33,angle=90.0]{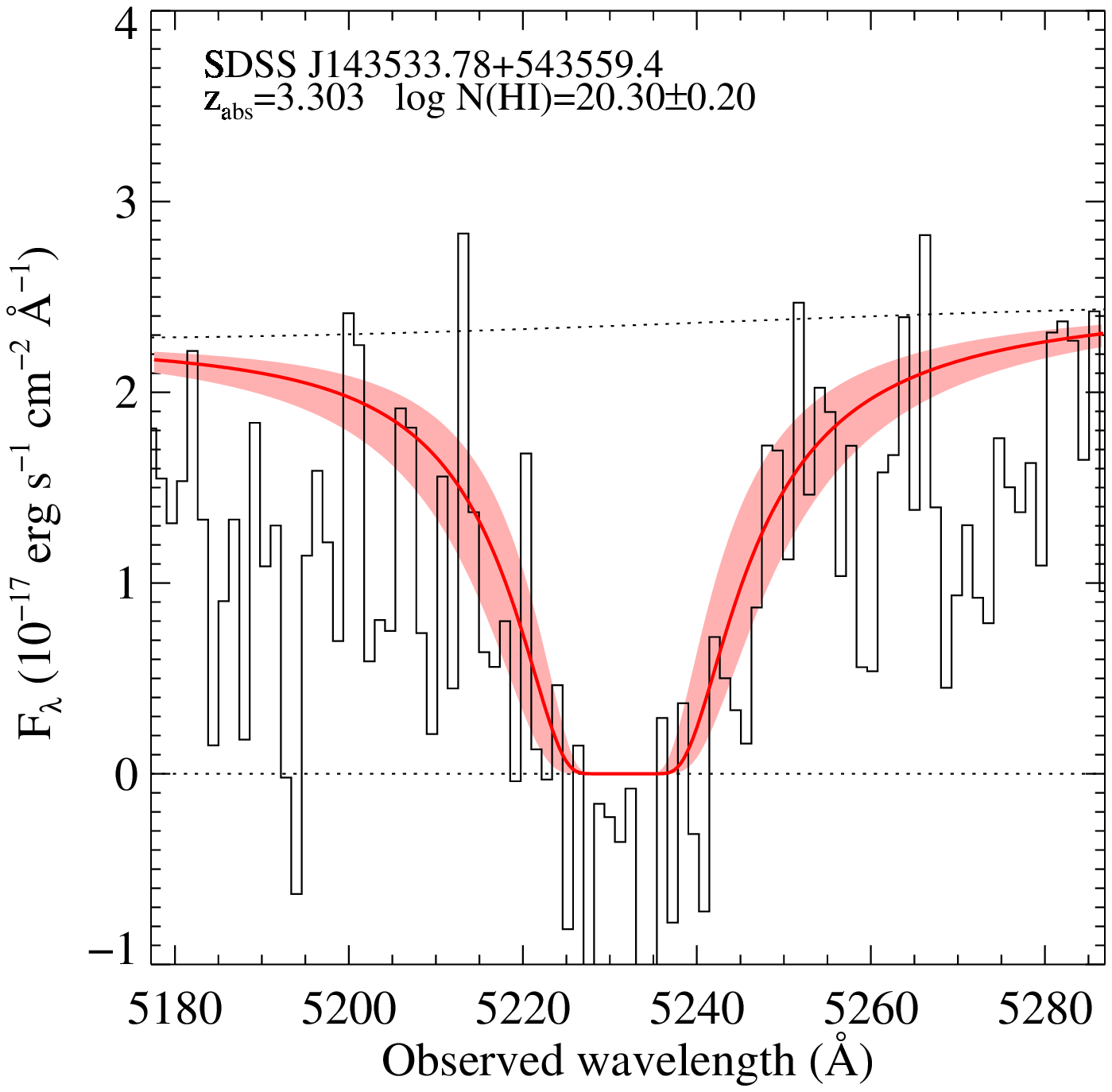}
}
}}
\caption[]{
SDSS spectra showing the \lya lines for 12 DLAs in our sample. 
The best fitted Voigt profiles (solid curves) together with the 
associated 1$\sigma$ errors (shaded regions) are over-plotted. The dotted
curve gives the best fitted continuum (in some cases the continuum fit
includes the emission lines also). We have used VLT UVES spectra
to get $N$(H~{\sc i}) in the case of \zabs = 3.1745 system towards J1337+3152
\citep[see][]{Srianand10}
and \zabs = 2.595 and  2.622 systems towards J0407$-$4410 \citep[CTS 247, see][]{Ledoux03}.
}
\label{dlafits}
\end{figure*}
This is why it is important to search for 21-cm absorption in DLAs  
over a wide redshift range. While a good fraction of DLAs/sub-DLAs 
preselected through
Mg~{\sc ii} absorption seems to show 21-cm absorption at $z\sim 1.3$
\citep[see for example][]{Gupta09, Kanekar09mg2},
searches for 21-cm absorption in DLAs at $z_{\rm abs}\ge2$ have
mostly resulted in null detections \citep[see][]{Kanekar03,Curran10}
with only four detections reported till now \citep[see][]{Wolfe85,Kanekar06,Kanekar07,York07}. 
{
The low detection rate of 21-cm absorption in high-$z$ DLAs can be
related to either the gas being warm (i.e high spin temperature, $T_{\rm S}$, 
as suggested by \citet{Kanekar03}) 
and/or the low value of covering factor ($f_c$)
through high-z geometric effects
\citep{Curran06}. 

The best way to address the covering factor issue is to perform 
milliarcsecond scale spectroscopy in the redshifted 21-cm line 
using very long baseline interferometry (VLBI) to measure the extent 
of absorbing gas \citep{Lane00}. Unfortunately due to limited 
frequency coverage and sensitivity of the receivers available with 
VLBI such studies cannot be extended to high redshift DLAs.  
Alternatively, the core fraction measured in the  milliarcsecond scale
images can be used to get an estimate of the covering factor
\citep[see][] {Briggs89, Kanekar09vlba}. Here one assumes that the 
absorbing gas completely covers at least the emission from the milliarcsec
scale core.
Therefore, to address this issue, 
one needs, not only to increase the number of systems searched for 
21-cm absorption but also to perform milliarcsecond scale imaging of
the background radio sources.

We report here the results of a search for 21-cm  absorption in
10 DLAs at $z>2$ we have carried out using GBT and GMRT, 
complemented by L-band VLBA images of the background QSOs. 
\begin{table}
\caption{Log of GBT and GMRT observations to search for 21-cm absorption}
\begin{center}
\begin{tabular}{lccccc}
\hline
\hline
\multicolumn{1}{c}{Source}& Tele-    & Date       & Time    &  BW      &    Ch.         \\
\multicolumn{1}{c}{name}  & scope    &            &         &          &                \\
            &          & yy-mm-dd           & (hr)    & (MHz)    &                \\
\multicolumn{1}{c}{(1)} & (2)      &  (3)       &  (4)    &   (5)    &    (6)         \\
\hline

J0407$-$4410 &  GBT  &  2006-10-20  &   4.7   &  0.625   &    512         \\  
(CTS247)  &       &  2007-01-05  &         &          &                \\  
             &       &  2007-01-06  &         &          &                \\  
             &       &  2007-01-08  &         &          &                \\  
 J0733+2721  &  GBT  &  2007-12-05  &  10.7   &  1.25    &    512         \\  
             &       &  2007-12-06  &         &          &                \\  
 J0801+4725  &  GMRT &  2006-12-22  &  10.8   &  1       &    128         \\  
             &       &  2006-11-23  &         &          &                \\
 J0852+2431  &  GBT  &  2009-08-07  &   5.6   &  1.25    &    1024        \\  
             &       &  2009-08-08  &         &          &                \\  
             &       &  2009-09-06  &         &          &                \\  
 J1017+6116  &  GBT  &  2008-10-15  &   4.5   &  1.25    &    1024        \\  
             &       &  2008-10-16  &         &          &                \\  
             &       &  2008-10-18  &         &          &                \\
 J1242+3720  &  GMRT &  2007-06-08  &   6.1   &  0.5     &    128         \\  
             &       &  2007-06-10  &         &          &                \\  
 J1337+3152  &  GMRT &  2009-01-13  &   6.2   &  1       &    128         \\  
             &       &  2009-03-17  &   7.8   &  0.25    &    128         \\  
 J1406+3433  &  GBT  &  2009-03-05  &   8.0   &  1.25    &    1024        \\  
             &       &  2009-03-06  &         &          &                \\  
             &       &  2009-03-07  &         &          &                \\  
             &       &  2009-03-08  &         &          &                \\  
             &       &  2009-04-07  &         &          &                \\  
 J1435+5435  &  GMRT &  2007-06-10  &   6.7   &  1       &    128         \\  
             &       &                 &         &          &                \\  
\hline
\end{tabular}
\end{center}
\begin{flushleft}
Column 1: Source name. 
Column 2: Telescope used for 21-cm absorption search. 
Column 3: Date of observations. 
Column 4: Total time on source i.e. after excluding telescope set-up time and calibration overheads. 
Columns 5 and 6: Spectral setup for the observations i.e. bandwidth (BW) and number of 
spectral channels respectively.  
\end{flushleft}
\label{obslog}
\end{table}
%
This survey has resulted in the detection of 21-cm absorption in 
the \zabs = 3.1745
DLA towards J1337+3152. A detailed analysis of this system and
two sub-DLAs close to this system are presented in \citet{Srianand10}.
Section~\ref{samp} presents the details of our sample. 
In Section 3 we present the details of  GBT and GMRT spectroscopic 
observations, VLBA continuum observations, and data reduction. 
The detection rate of 21-cm absorption in DLAs is discussed in 
Section~\ref{detect}. In Section~\ref{gencor} we study the
correlations between the parameters derived from 21-cm observations, 
$N$(H~{\sc i}), metallicity
and redshift. In Section~\ref{mole} we study the relation between
21-cm and \h2 absorption. The results  are summarized in 
Section~\ref{results}. In this work we assume a flat Universe with
$H_0$~=~71~\kms~Mpc$^{-1}$, $\Omega_{\rm m}$~=~0.27 and $\Omega_\Lambda$~=~0.73.

\section{The sample of DLAs}
\label{samp}

\begin{table*}
\caption{GMRT low-frequency flux density measurements for the DLAs observed with the GBT 
}
\begin{center}
\begin{tabular}{lcccc}
\hline
\hline
\multicolumn{1}{c}{Source name}       & $S_{\rm 610MHz}$ &   Date              & $S_{\rm 325MHz}$ &    Date          \\
         &                &                       &                &                  \\
             &  (mJy)         &   yy-mm-dd       &   (mJy)        &   yy-mm-dd               \\
\multicolumn{1}{c}{(1)}  &   (2)          &    (3)                &    (4)         &     (5)          \\
\hline                                                                                  
J0407$-$4410 &  124           &  2006-05-23        &    52          & 2007-11-24     \\
J0733$+$2721 &  248           &  2007-11-04        &    549         &      ''          \\
J0852$+$2431 &  198           &  2008-12-26        &    237         & 2009-03-17     \\
J1017$+$6116 &  274           &     ''                &    266         &      ,,          \\
J1406$+$3433 &  165           &     ''                &    185$^\dag$  &     (WENSS)      \\
\hline
\end{tabular}
\end{center}
\begin{flushleft}
Column 1: Source name.  
Columns 2 and 4: GMRT flux density measurements at 610 and 325\,MHz respectively. 
Columns 3 and 5: Dates of the 610 and 325\,MHz observations respectively. \\
$^\dag$ From the WENSS catalog.\\ 
\end{flushleft}
\label{flux}
\end{table*}
\begin{table*}
\caption{Results from the GBT and GMRT observations}
\begin{center}
\begin{tabular}{lccccccccc}
\hline
\hline
Source name &   \zem   & \zabs  &  log\,$N$(H~{\sc i})   &S$_{1.4\,GHz}$& $\delta $  &Spectral rms & S$_{\nu_{abs}}$ & $\int\tau$dv  & {${ T}_{\rm s}\over f_{\rm c}$}   \\
            &          &        &        (cm$^{-2}$)    &   (mJy)      & (km\,s$^{-1}$) &(mJy\,b$^{-1}$\,ch$^{-1}$)  & (mJy)&  (km\,s$^{-1}$)&(K)  \\
~~~~~~~~(1) & (2)      &  (3)   &  (4)            &   (5)        &     (6)          &    (7)        &    ~~(8)     &  (9)  & (10) \\
\hline
\\
 J040718$-$441013    & 3.020 &2.595  & 21.05$\pm$0.10 &   -     & 3.7   & 5.9    &   67  & $<$1.61  &  $>$382    \\  
 J040718$-$441013    &  ''   &2.622  & 20.45$\pm$0.10 &   -     & ''    & 7.1    &   67  & $<$1.93  &  $>$81     \\  
 J073320.49+272103.5 & 2.938 &2.7263 & 20.25$\pm$0.20 &  240    & 3.8   & 3.4    &  451  & $<$0.14  &  $>$692    \\  
 J080137.68+472528.2 & 3.276 &3.2235 & 20.80$\pm$0.15 &   78    & 7.0   & 1.5    &  164  & $<$0.22  &  $>$1563   \\  
 J085257.12+243103.2 & 3.617 &2.7902 & 20.70$\pm$0.20 &  160    & 3.9   & 3.9    &  228  & $<$0.32  &  $>$850    \\  
 J101725.89+611627.5 & 2.805 &2.7681 & 20.60$\pm$0.15 &  477    & 3.9   & 4.2    &  268  & $<$0.29  &  $>$758    \\  
 J124209.81+372005.7 & 3.839 &3.4135 & 20.50$\pm$0.30 &  662    & 3.6   & 3.6    &  615  & $<$0.11  &  $>$1567   \\  
 J133724.69+315254.5 & 3.174 &3.1745 & 21.36$\pm$0.10 &   83    & 6.9   & 1.3    &   69  &2.08$\pm$0.17 &  600$^{+220}_{-160}$\\  
 J140653.84+343337.4 & 2.566 &2.4989 & 20.20$\pm$0.20 &  167    & 3.6   & 3.0    &  178  & $<$0.31  & $>$356    \\  
 J143533.78+543559.4 & 3.811 &3.3032 & 20.30$\pm$0.20 &   96    & 7.1   & 1.5    &  145  & $<$0.26  & $>$418    \\  
                     &       &       &                &         &       &        &       &          &           \\  
\hline
\end{tabular}
\end{center}
\begin{flushleft}
Column 1: Source name. 
Column 2: QSO emission redshift. 
Column 3: Absorption redshift of DLAs as determined from the metal absorption lines. 
Column 4: H~{\sc i} column density.  
Column 5: Flux density at 1.4\,GHz from the FIRST catalog.
Columns 6 and 7: Spectral resolution and rms for the survey spectrum.  
Column 8: Continuum flux density of source at the redshifted 21-cm absorption frequency. 
Column 9: Integrated 21-cm optical depth or 3-sigma upper limit to $\int \tau dv$ for the equivalent spectral resolution of 
10\,km\,s$^{-1}$. 
Column 10: Ratio of spin temperature and covering factor of absorbing gas. 
\\
\end{flushleft}
\label{dlasamp}
\end{table*}
%
To construct our sample, we cross-correlated the overall sample of 
DLA-bearing QSO sightlines from SDSS-DR7 
\citep[][including systems that are not part
of the published statistical sample used to measure $\Omega_{\rm HI}$]{Noterdaeme09dla} with the VLA FIRST catalog to
identify DLAs in front of radio sources brighter than 50~mJy at 1.4 GHz.  
We excluded radio sources with the DLA 21-cm absorption frequencies redshifted 
into GBT and GMRT frequency ranges known to be affected by 
strong radio frequency interference (RFI). There are 13 DLAs from the SDSS-DLA catalog that satisfy these conditions.  
In addition, there are 4 DLAs along the sight line towards J0407$-$4410
(also known as CTS\,247) at \zabs~=~1.913, 2.550, 2.595 \& 2.622. Two of these,
at \zabs~=~2.595 and 2.622, have redshifted 21-cm absorption frequency in the 
relatively RFI free frequency range of GBT. Including these two DLAs towards 
CTS\,247, we have a sample of 15 DLAs for which a search
for 21-cm absorption was carried-out using either GMRT or GBT. We observed 14 DLAs, (the exception is the \zabs = 3.079 system towards 
J1413+4505), but obtained useful spectra for only 10 DLAs.
In addition, we have obtained milliarcsecond scale 
images at 1.4\,GHz for all the QSOs except CTS\,247 to understand the role of 
radio structure in detectability of 21-cm absorption in DLAs. 
The details of the GBT, GMRT and VLBA observations are given below.  

The Lyman-$\alpha$ profiles for 12 DLAs selected from the SDSS-DLA catalog 
are shown in the Fig.~\ref{dlafits}.  The H~{\sc i} column density for 
each of these DLAs has been estimated using Voigt profile fits to the 
Lyman-$\alpha$ absorption line.  The QSO continuum was approximated by a 
lower order spline using absorption free regions on both sides of 
the H~{\sc i} trough (dotted curves in each panel). In addition, 
special care was taken to fit the emission line profiles whenever 
the \lya absorption is close to QSO emission lines. 
For the remaining three DLAs in our sample, \zabs=3.1745 DLA towards 
J1337+3152 and the two DLAs towards CTS\,247, we use the column 
densities measured by 
\citet{Srianand10} and \citet{Ledoux03} respectively from high 
resolution VLT UVES spectra.

%

\section{Details of Observations and data reduction}
\subsection{The GBT and GMRT observations}

We observed our sample of 14 DLAs using the GBT prime focus receivers 
PF1-340\,MHz and PF1-450\,MHz, and the GMRT P-band receiver.  Although 
we selected DLAs such that the redshifted 21-cm absorption frequencies were 
not affected by strong RFI, no useful data could be obtained for 
4 absorption systems either due to 
RFI or other technical reasons.  The observing log for the remaining 
10 DLAs and the spectral set-up used for these observations are 
provided in Table~\ref{obslog}.  
GBT observations were performed in the standard position-switching mode with 
typically 5\,min spent on-source and 5\,min spent off-source. 
The data were acquired in the orthogonal polarization 
channels XX and YY. We used the 
GBT spectral processor as the backend for these observations.  
The two DLAs towards 
CTS\,247 were observed simultaneously using two bands of 0.625\,MHz split 
into 512 channels.
For the GMRT observations, typically a  
bandwidth of 0.5 or 1\,MHz 
split into 128 frequency channels was used. The data were acquired in the two 
orthogonal polarization channels RR and LL.  
For the flux density/bandpass calibration of GMRT 
data, standard flux density calibrators were observed for 10-15\,min every 
two hours.  A phase calibrator was also observed for 10\,min every 
$\sim$45\,min to get reliable phase solutions.    

We used NRAO's GBTIDL package to develop a pipeline to automatically 
analyse the GBT spectral-line data sets. After excluding time ranges for 
which no useful data were obtained, the data were processed through this 
pipeline. The pipeline calibrates each data record individually and flags 
the spectral channels with deviations larger than 5$\sigma$ as 
affected by RFI. After subtracting a second order baseline these data 
are averaged to produce baseline (i.e. continuum) subtracted spectra for 
XX and YY. The baseline fit and statistics 
for the flagging are determined using the spectral region that excludes the
central 25\% and last 
10\% channels at both ends of the spectrum. If necessary, a first-order cubic spline was 
fitted to the averaged XX and YY spectra obtained from the pipeline, which were then combined 
to produce the Stokes-I spectrum. The spectrum was then shifted to the heliocentric frame.  
The multi-epoch spectra for a source were then resampled onto the same frequency scale and 
combined to produce the final spectrum.

The GMRT data were reduced using the NRAO AIPS package 
following the standard procedures  described in \citet{Gupta06}.  
Special care was taken to exclude the baselines and time stamps affected 
by RFI.  
The spectra at the quasar positions were extracted from the RR and LL 
spectral cubes and compared for consistency. If necessary, a first-order 
cubic-spline was fitted to remove the residual continuum from the spectra.  
The two polarization channels were then combined to get the stokes I spectrum 
which was then shifted to the heliocentric frame.

\begin{figure*}
\centerline{\hbox{
\psfig{figure="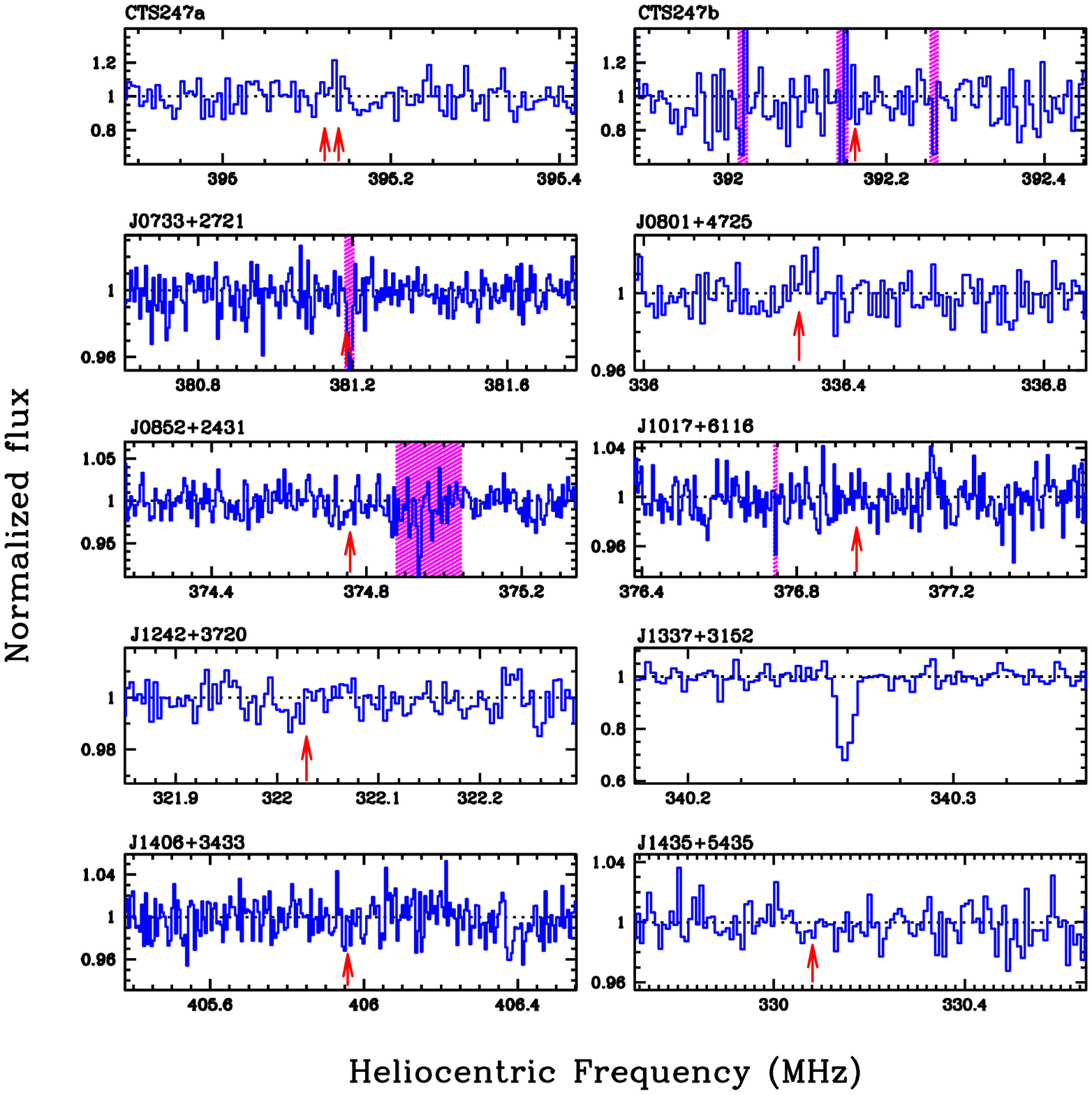",height=16.0cm,width=17.0cm,angle=0}
}}
\caption[]{GBT and GMRT spectra of DLAs in our sample. 
Shaded regions mark features that are due to RFI. The arrows in the case of non-detections indicate the expected
positions of the 21-cm absorption. In the case of \zabs = 3.1745 system towards
J1337+3152 we show the high resolution 21-cm absorption spectrum only. Two arrows in the case of
CTS247a indicate the expected position of 21-cm absorption from the two H$_2$ components.
}
\label{21cm}
\end{figure*}
The FWHM of the GBT beam at 400\,MHz is 30$^\prime$ and the rms confusion is 
500\,mJy. This is comparable  
to the flux densities of the background radio sources observed with the GBT. 
{Therefore, to  
correct for the effect of other confusing sources in the GBT beam and determine 
the QSO flux densities at the redshifted 21-cm frequency, we observed these with the GMRT at 610 and 325\,MHz. For these observations we 
have used 32\,MHz bandwidth.  Details of these GMRT observations 
and the measured flux densities are provided in Table~\ref{flux}.}  For J1406+3433, the 325\,MHz 
flux density is taken from the Westerbork Northern Sky Survey (WENSS). We interpolate these flux density measurements to determine 
the flux densities at redshifted 21-cm frequencies for the quasars observed with the GBT.  
Since, flux densities for these 5 QSOs are not measured at the same epoch as the 
GBT spectroscopic observations, in principle, radio flux density 
variability can affect 
our estimates of 21-cm optical depths for the corresponding DLAs.  However, 
this effect is much smaller than the error caused by confusion from other sources in the beam and should not 
affect the statistical results derived later in the paper.  

%
The GBT and GMRT observations of our DLA sample have resulted in useful 
21-cm absorption spectra for 10 DLAs.  These spectra are presented 
in Fig.~\ref{21cm}. GBT spectra, typically acquired at a spectral 
resolution, of $\sim$1\,\kms 
have been smoothed to $\sim$4\,\kms for presentation. 
The 21-cm absorption is detected only for one DLA (i.e. \zabs~=~3.1745 DLA towards J1337+3152)
and a detailed analysis of this system is presented in \citet{Srianand10}.
None of the other ``absorption-like features'', marked as shaded regions, 
are reproduced in  spectra from different polarizations and epochs, 
but are due to RFI.  For CTS247b (i.e for \zabs = 2.622 DLA towards
CTS247)  these 
features are present only in one polarization at certain times.  For J0852+2431  
and J1017+6116, using a combination of high spectral resolution ($\sim$1\,\kms) and/or 
multi-epoch observations we rule out the possibility of these features 
being real 21-cm absorption.  
Details of the optical depth measurements and other observational results for all the 
10 DLAs are summarized in Table~\ref{dlasamp}.

\begin{table}
\caption{Details of phase-referencing calibrators used for the VLBA observations }
\begin{center}
\begin{tabular}{lcc}
\hline
\hline
Source      & Calibrator    & Separation        \\
            &               &(degrees)\\
~~~~~~~~(1) & (2)           &  (3)              \\
\hline
 J0733+2721 & J0732+2548    &   1.5             \\  
 J0801+4725 & J0754+4823    &   1.5             \\  
 J0816+4823 & J0808+4950    &   1.9             \\
 J0839+2002 & J0842+1835    &   1.6             \\
 J0852+2431 & J0856+2111    &   3.5             \\  
 J1017+6116 & J1031+6020    &   2.0             \\  
 J1223+5037 & J1227+4932    &   1.3             \\
 J1237+4708 & J1234+4753    &   0.9             \\
 J1242+3720 & J1242+3751    &   0.5             \\  
 J1337+3152 & J1329+3154    &   1.6             \\  
 J1406+3433 & J1416+3444    &   1.9             \\  
 J1413+4505 & J1417+4607    &   1.2             \\
 J1435+5435 & J1429+5406    &   1.0             \\  
\hline
\end{tabular}
\end{center}
\begin{flushleft}
Column 1: Source name. 
Column 2: Phase-referencing calibrator. 
Column 3: Separation between the radio source and phase-referencing calibrator. 
\end{flushleft}
\label{vlbalog}
\end{table}

\subsection{Continuum observations with VLBA}
The sample of quasars presented here was observed as part of a 
larger VLBA survey to obtain milliarcsecond scale images for QSOs 
with foreground DLAs and Mg~{\sc ii} systems,  
and understand the relationship between radio structure 
and detectability of 21-cm 
absorption.   We have observed using VLBA 21-cm receiver for 11~hrs and  18~hrs
on 21/02/2010 and 10/06/2010 respectively.
We used eight 8\,MHz baseband channels, i.e. the total bandwidth of 64\,MHz. 
Each baseband channel was split into 32 spectral points.  Both the right 
and left-hand circular polarization channels were recorded.  Two bit 
sampling and a post-correlation time resolution of 2 seconds were used. 

The observations were done using nodding-style phase-referencing with a cycle 
time of $\sim$5\,min, i.e. 3\,min on the source and $\sim$1.5\,min on the 
phase-referencing calibrator. The phase-referencing calibrators were selected 
from the VLBA calibrator survey (VCS) at 2.3 and 8.6\,GHz 
(Table~\ref{vlbalog}).  
In order to improve the uv-coverage, the total observing time was split 
into snapshots over a number of different hour angles. Each source, except 
CTS247 which was excluded due to observational constraints, was typically 
observed for a total of $\sim$30\,min. During both observing runs, strong 
fringe finders/bandpass calibrators such as J0555+3948, J0927+3902, 
J1800+3848 and J2253+1608 were also observed every $\sim$3\,hr for 4-5\,min. 
\begin{figure*}
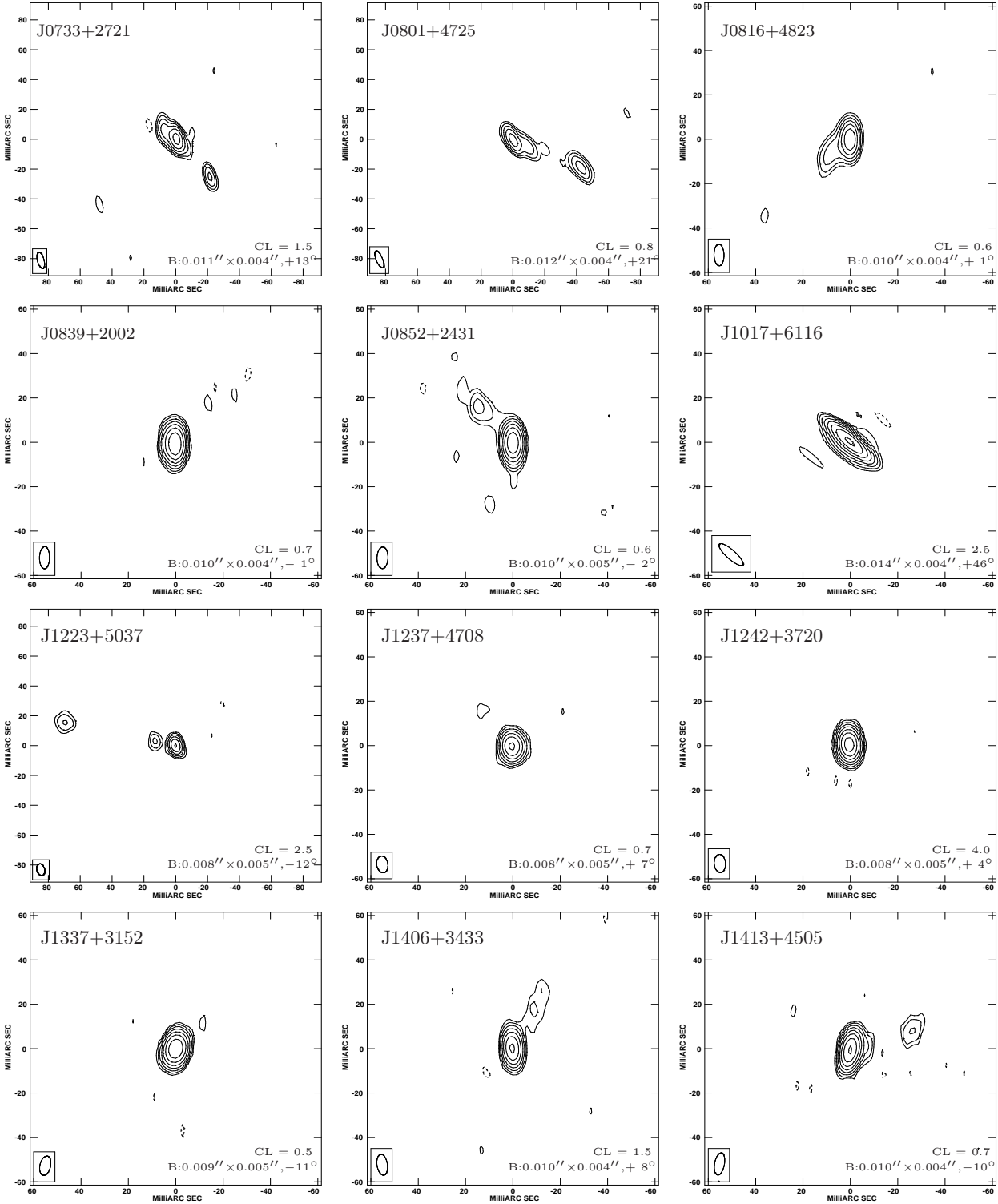

\centerline{
\vbox{
\hbox{
\psfig{figure="J0733MAP_NOLABELS.PS",height=5.0cm,width=5.5cm,angle=-90}
\psfig{figure="J0801MAP_NOLABELS.PS",height=5.0cm,width=5.5cm,angle=-90}
\psfig{figure="J0816MAP_NOLABELS.PS",height=5.0cm,width=5.5cm,angle=-90}
}
\hbox{
\psfig{figure="J0839MAP_NOLABELS.PS",height=5.0cm,width=5.5cm,angle=-90}
\psfig{figure="J0852MAP_NOLABELS.PS",height=5.0cm,width=5.5cm,angle=-90}
\psfig{figure="J1017MAP_NOLABELS.PS",height=5.0cm,width=5.5cm,angle=-90}
}
\hbox{
\psfig{figure="J1223MAP_NOLABELS.PS",height=5.0cm,width=5.5cm,angle=-90}
\psfig{figure="J1237MAP_NOLABELS.PS",height=5.0cm,width=5.5cm,angle=-90}
\psfig{figure="J1242MAP_NOLABELS.PS",height=5.0cm,width=5.5cm,angle=-90}
}
\hbox{
\psfig{figure="J1337MAP_NOLABELS.PS",height=5.0cm,width=5.5cm,angle=-90}
\psfig{figure="J1406MAP_NOLABELS.PS",height=5.0cm,width=5.5cm,angle=-90}
\psfig{figure="J1413MAP_NOLABELS.PS",height=5.0cm,width=5.5cm,angle=-90}
}
}
}
\caption[]{Contour plots of VLBA images at 1.4\,GHz. The rms in the images are listed 
in Table~\ref{vlbares}. 
{At the bottom of each image the restoring beam is shown as an ellipse, 
and the first contour level (CL) in mJy\,beam$^{-1}$ and FWHM are noted}.  
The contour levels are plotted as CL$\times$($-$1, 1, 2, 4, 8,...)\,mJy\,beam$^{-1}$. {Depending upon the detailed structure of the radio sources,
the emission could be more extended at the redshifted
21-cm frequencies.}
}
\vskip -22.3cm
\begin{picture}(400,400)(0,0)
\put(-019,373){\small      J0733$+$2721}
\put( 084,272){\tiny       CL = 1.5}
\put( 043,265){\tiny       B:0.011$^{\prime\prime}$$\times$0.004$^{\prime\prime}$,$+$13$^\circ$ }
\put( 144,373){\small      J0801$+$4725}
\put( 245,272){\tiny       CL = 0.8}
\put( 205,265){\tiny       B:0.012$^{\prime\prime}$$\times$0.004$^{\prime\prime}$,$+$21$^\circ$ }
\put( 304,373){\small      J0816$+$4823}
\put( 405,272){\tiny       CL = 0.6}
\put( 364,265){\tiny       B:0.010$^{\prime\prime}$$\times$0.004$^{\prime\prime}$,$+$ 1$^\circ$ }
\put(-017,230){\small      J0839$+$2002}
\put( 084,129){\tiny       CL = 0.7}
\put( 043,122){\tiny       B:0.010$^{\prime\prime}$$\times$0.004$^{\prime\prime}$,$-$ 1$^\circ$ }
\put( 144,230){\small      J0852$+$2431}
\put( 244,129){\tiny       CL = 0.6}
\put( 204,122){\tiny       B:0.010$^{\prime\prime}$$\times$0.005$^{\prime\prime}$,$-$ 2$^\circ$ }
\put( 304,230){\normalsize J1017$+$6116}
\put( 404,129){\tiny       CL = 2.5}
\put( 363,122){\tiny       B:0.014$^{\prime\prime}$$\times$0.004$^{\prime\prime}$,$+$46$^\circ$ }
\put(-017,087){\normalsize J1223+5037}
\put( 084,-13){\tiny       CL = 2.5}
\put( 043,-20){\tiny       B:0.008$^{\prime\prime}$$\times$0.005$^{\prime\prime}$,$-$12$^\circ$ }
\put( 144,087){\normalsize J1237$+$4708}
\put( 244,-13){\tiny       CL = 0.7}
\put( 204,-20){\tiny       B:0.008$^{\prime\prime}$$\times$0.005$^{\prime\prime}$,$+$ 7$^\circ$ }
\put( 304,087){\normalsize J1242$+$3720}
\put( 404,-13){\tiny       CL = 4.0}
\put( 363,-20){\tiny       B:0.008$^{\prime\prime}$$\times$0.005$^{\prime\prime}$,$+$ 4$^\circ$ }

\put(-017,-56){\normalsize J1337$+$3152}
\put( 084,-156){\tiny      CL = 0.5}
\put( 043,-163){\tiny      B:0.009$^{\prime\prime}$$\times$0.005$^{\prime\prime}$,$-$11$^\circ$ }
\put( 144,-56){\normalsize J1406$+$3433}
\put( 244,-156){\tiny      CL = 1.5}
\put( 204,-163){\tiny      B:0.010$^{\prime\prime}$$\times$0.004$^{\prime\prime}$,$+$ 8$^\circ$ }
\put( 304,-56){\normalsize J1413$+$4505}
\put( 404,-156){\tiny      CL = 0.7}
\put( 363,-163){\tiny      B:0.010$^{\prime\prime}$$\times$0.004$^{\prime\prime}$,$-$10$^\circ$ }
\end{picture}
\vskip +9.5cm
\label{vlbamap}
\end{figure*}
\begin{figure}
\addtocounter{figure}{-1}
\centerline{
\vbox{
\hbox{
\psfig{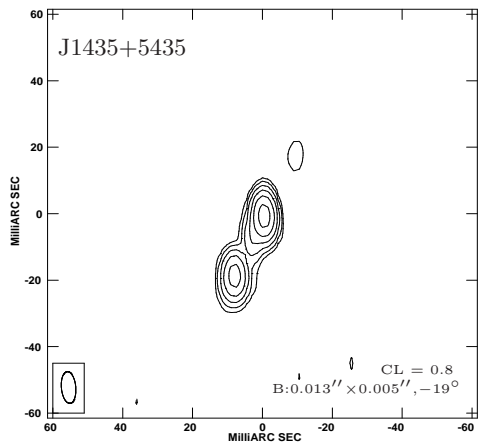}
}
}
}
\caption[]{ {\sl Continued}. 
}
\vskip -7.1cm
\begin{picture}(400,400)(0,0)
\put( 050,373){\normalsize J1435$+$5435}
\put( 172,252){\tiny       CL = 0.8}
\put( 131,245){\tiny       B:0.013$^{\prime\prime}$$\times$0.005$^{\prime\prime}$,$-$19$^\circ$ }
\end{picture}
\vskip -7.1cm
\label{vlbamap}
\end{figure}

\begin{sidewaystable*}[]
\vskip -7 in
%
\vskip 0.3in
\centering
\begin{tabular}{lcccccccccccccc}
\multicolumn{15}{l}{{\bf Table 5}: ~Results from the VLBA data. }
\\
\hline
\hline
Source      &$z_{\rm abs}$&  Right       &   Declination       & rms &  Comp. &  S  &  r  &$\theta$&  a & b/a &$\phi$& $S_{\rm T}$ & $f_{\rm VLBA}$ &LLS \\
name        &       &  ascension     &                     &     &        &     &     &        &    &     &      &       &       &    \\
            &       &   (J2000)      &   (J2000)&(mJy\,beam$^{-1}$)&        &(mJy)&(mas)& ($^\circ$)&(mas)& &($^\circ$)& (mJy) &       &(pc)\\
 (1)        &  (2)  &      (3)       &       (4)           & (5) &  (6)   & (7) & (8) & (9)    &(10)&(11) &(12)  & (13)  &  (14) &(15)\\
\hline
J0733+2721  & 2.7263& 07 33 20.4830 & +27 21 03.430  & 0.3 &   1    & 97  &  0  &   -    &4.52 &0.05 &-88   & 240   & 0.62  &348 \\%
            &       &               &                &     &   2    & 3   & 9.1 &-129    &13.24&0.00 &-18   &       &       &   \\%
            &       &               &                &     &   3    & 19  &32.7 &-140    &5.98 &0.26 & 32   &       &       &   \\%
            &       &               &                &     &   4    & 29  &10.8 &  51    &8.37 &0.44 &-17   &       &       &   \\%
J0801+4725  & 3.2235& 08 01 37.6930 & +47 25 28.082  & 0.2 &   1    & 28  &  0  &   -    &6.77 &0.00 & 72   & 78    & 0.67  &342 \\%
            &       &               &                &     &   2    & 8   &9.34 &-112    &7.09 &0.00 & 72   &       &       &   \\%
            &       &               &                &     &   3    & 16  &44.7 &-112    &8.63 &0.22 & 64   &       &       &   \\%
J0816+4823  &       & 08 16 19.0044 & +48 23 28.490  & 0.2 &   1    & 42  &  0  &   -    &4.59 &0.24 & 87   & 69    & 0.75  &68   \\%
            &       &               &                &     &   2    & 10  & 9.0 &  132   &8.75 &0.43 &-63   &       &       &    \\%
J0839+2002  &       & 08 39 10.8970 & +20 02 07.391  & 0.2 &   1    & 113 &  0  &   -    &4.91 &0.24 & 84   & 130   & 0.87  &$\le$39\\%
J0852+2431  & 2.7902& 08 52 57.1211 & +24 31 03.271  & 0.2 &   1    & 78  &  0  &   -    &4.92 &0.31 & 71   & 160   & 0.55  & 175\\%
            &       &               &                &     &   2    & 10  & 21.9&  42    &15.2 &0.16 & 53   &       &       &   \\%
J1017+6116  &2.7681 & 10 17 25.8865 & +61 16 27.414  & 0.5 &   1    &388  &  0  &   -    &1.50 &0.68 & 55   & 477   & 0.86  &38\\%
            &       &               &                &     &   2    & 24  & 4.7 &  145   &2.45 &0.00 & 70   &       &       &   \\%
J1223+5037  &       & 12 23 43.1740 & +50 37 53.344  & 0.5 &   1    & 96  &  0  &   -    &2.50 &0.00 & 78   & 229   & 0.60  &554   \\%
            &       &               &                &     &   2    & 16  &13.6 &  80    &3.01 &0.83 &-19   &       &       &   \\%
            &       &               &                &     &   3    & 25  &71.4 &  78    &8.61 &0.00 & 89   &       &       &   \\%
J1237+4708  &       & 12 37 17.4413 & +47 08 06.964  & 0.2 &   1    & 64  &  0  &   -    &3.17 &0.20 &-74   & 80    & 0.80  & $\le27$\\%
J1242+3720  &3.4135 & 12 42 09.8121 & +37 20 05.692  & 0.6 &   1    & 848 &  0  &   -    &1.93 &0.76 & 22   & 662   & 1.00  & $\le14$  \\%
J1337+3152  &3.1747 & 13 37 24.6931 & +31 52 54.642  & 0.2 &   1    & 83  &  0  &   -    &3.85 &0.38 & 74   & 83    & 1.00  &$\le30$   \\%
J1406+3433  &2.4989 & 14 06 53.8532 & +34 33 37.339  & 0.4 &   1    &127  &  0  &   -    &3.24 &0.22 &-23   & 167   & 0.87  & 153  \\%
            &       &               &                &     &   2    & 18  & 18.7& -30    &23.79&0.23 &-25   &       &       &   \\%
J1413+4505  &       & 14 13 18.8652 & +45 05 22.990  & 0.2 &   1    & 105 &  0  &   -    &2.19 &0.42 &-67   & 140   & 0.88  & 216  \\%
            &       &               &                &     &   2    & 12  & 3.4 & -77    &5.37 &0.00 &-86   &       &       &   \\%
            &       &               &                &     &   3    & 6   &27.9 & -72    &6.28 &0.03 &-67   &       &       &   \\%
J1435+5435  &3.3032 & 14 35 33.7812 & +54 35 59.312  & 0.2 &   1    & 31  &  0  &   -    &2.66 &0.17 &-29   & 96    & 0.55  &155   \\%
            &       &                  &                   &     &   2    & 17  &20.4 &  155   &4.00 &0.47 &-36   &       &       &   \\%
            &       &                  &                   &     &   3    & 5   & 7.6 &  153   &5.52 &0.00 &-7    &       &       &   \\%
            &       &                  &                   &     &        &     &     &        &     &     &      &       &       &   \\%
\hline
\end{tabular}
\begin{flushleft}
Column 1:~Source name. Column 2:~absorption redshift. 
Columns 3 and 4:~right ascension and declination of component-1 (see column 6)
from the multiple Gaussian fit to the source, respectively. 
Column 5:~rms in the map in mJy\,beam$^{-1}$. 
Column 6:~component id. 
Column 7:~flux density of the component in mJy.
Columns 8 and 9:~radius and position angle of the component with respect to component-1, respectively.
{Columns 10, 11 and 12: major axis, axial ratio and position angle 
of the deconvolved Gaussian component, respectively.}
Column 13:~flux density in mJy from FIRST/NVSS. 
Column 14:~$c_f$ is the ratio of 1.4 GHz flux density in 
VLBA image to that in the FIRST image, 
Column 15:~largest projected linear size in pc. \\
\end{flushleft}
\label{vlbares}
\end{sidewaystable*}

Data were calibrated and imaged using AIPS and DIFMAP in a standard way. 
Global fringe fitting was performed on the phase-referencing calibrators.  
The delays, rates and phases estimated from these were transferred to the 
sources which were then self-calibrated until the final images were obtained. 
Radio sources were characterised by fitting Gaussian models to the self-calibrated 
visibilities. 
VLBA maps of the 13 QSOs are shown in Fig.~\ref{vlbamap} 
and the results of model fitting are listed in columns\,\#\,6-12 of Table~\ref{vlbares}.  

Non-detection of 21-cm absorption in a DLA could be due to the small covering 
factor of the absorbing gas. The typical spatial resolution achieved in our 
VLBA observations is $\sim$8\,mas. 
If the extent of absorbing gas is of the order of the scales probed by our 
VLBA observations (i.e. $>$20pc) then we expect the detectability of 
21-cm absorption to depend on the fraction and spatial extent of radio 
flux density detected in these images.  In column\,\#\,14 of Table~\ref{vlbares} 
we give the ratio of total flux densities detected 
in the VLBA and FIRST images at 20cm, i.e. $f_{\rm VLBA}$. \
The last column of this table 
gives the largest linear size (LLS), i.e the separation between the farthest radio components, of the 
radio source at the redshift of the DLA.        
Out of the 13 QSOs presented in Fig.~\ref{vlbamap} that have DLAs along 
their line of sight, we have 21-cm absorption 
spectra for only 8 DLAs.  For the DLA towards 
J0816+4823, we use the 21-cm absorption measurement from   
\citet{Curran10}.  Thus we have a sample of 9 DLAs with both 21-cm 
absorption measurements and VLBA 21-cm maps for the background QSOs. 
The $f_{\rm VLBA}$ for this sample ranges from 0.6 to 1, and LLS from 
$<$15\,pc to 340\,pc.


In the absence of VLBI spectroscopy at the redshifted 21-cm line
frequency, the ratio of VLBA core flux density to the total flux density measured in the arcsecond scale
images (called core fraction $c_f$) has been used as an
indicator of the covering factor $f_c$ \citep[see][]
{Briggs89, Kanekar09vlba}. 
Here we use the term `core' to refer to the flat spectrum
unresolved radio component coincident with
the optical QSO in the VLBA image.

For J1242+3720 and J1337+3152 the radio source is modelled 
as a single unresolved component and within the uncertainties all 
the flux density in the arcsecond scale 
FIRST images is recovered in our VLBA images (Fig.~\ref{vlbamap}). 
Both of these sources,  have 
flat spectra
{suggesting the radio emission even at lower frequencies
originates predominantly from the compact core.
Therefore we take $f_c$=1 for this case.}

The radio source J1017+6116 has an inverted radio spectrum ($\alpha$=$-$0.4) 
 and 86\% of the 
flux density in the FIRST image is recovered in the VLBA image, 94\% of which is contained in the main unresolved
component (Fig.~\ref{vlbamap}).  
For another 3 flat-spectrum sources with 21-cm absorption measurements, i.e. 
J0816+4823, J0852+2431, and J1406+3433, more than 80\% of the VLBA flux 
density is present in a 
single unresolved component.    
For these 4 sources, based on the flat spectral index, the dominant
component in the VLBA image can be identified with radio core/optical
QSO.  
Therefore for the 6 sources mentioned above, we have estimated 
the core fraction, 
$c_f$, and used it as the covering factor, $f_c$, of the gas}. 
The remaining three sources, J0733+2721, J0801+4725 and J1435+5435, exhibit multiple components in their 
VLBA 20-cm images.  The identification of the component coincident with the 
QSO, 
and the estimation of their $c_f$ from VLBA images for these three
sources are highly uncertain.

\section{Detectability of 21-cm absorption}
\label{detect}
\begin{figure}
\centerline{
\includegraphics[scale=0.4,angle=0.0]{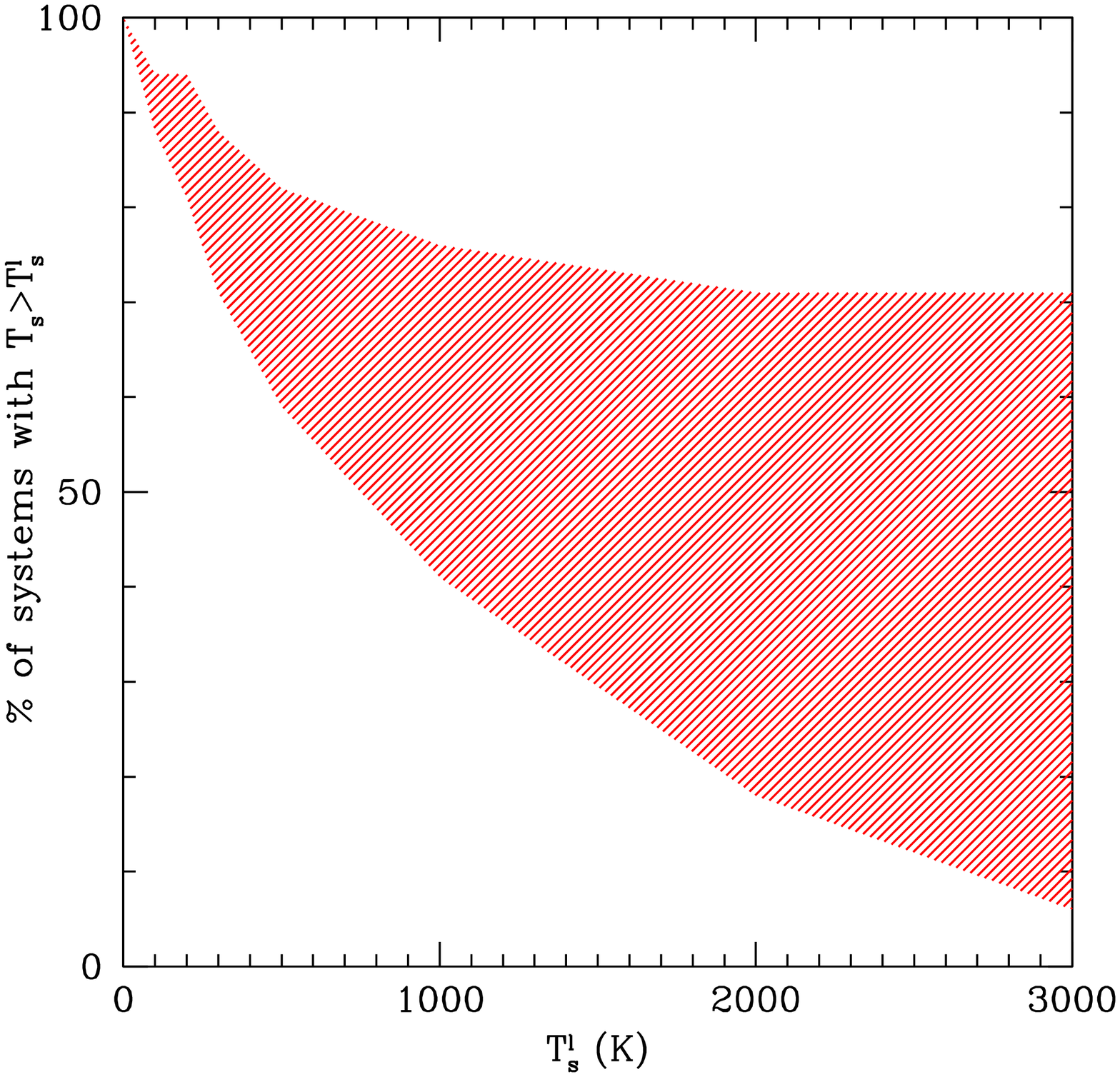}
}
\caption{ The allowed range of fraction of DLAs having harmonic mean
spin temperature $T_{\rm S}$ greater than a limiting value $T_{\rm S}^l$
as a function of $T_{\rm S}^l$. We use only those DLAs for which
$c_f$ measurements are available.  The lower envelope of the shaded 
region is obtained considering all the lower limits 
on $T_{\rm S}$  as measurements. The upper envelope is obtained 
assuming all the lower limits as measurements with 
$T_{\rm S}\ge T_{\rm S}^l$.
}
\label{fwnm}
\end{figure}

In this Section we investigate the detectability of 21-cm absorption in DLAs 
and the implication of non-detections for the physical state of the H~{\sc i} 
gas.
It is clear from the last column of Table~\ref{dlasamp}, 
that for most of the DLAs, our data has good sensitivity to detect 
$T_{\rm S}/f_{\rm c}$$\sim$100~K gas.

The H~{\sc i} 21-cm absorption is detected only in the \zabs~=~3.1745 system
towards SDSS~J1337+3152. {This is one of the weakest radio sources in
our sample (with a 3$\sigma$ $\int\tau dv$ limit of 0.4 \kms).  However, 
thanks to high $N$(H~{\sc i}) our spectrum is sensitive enough to detect
any gas with $T_{\rm S}\le 3100$ K.
}
 This source is unresolved in our
VLBA observations (see Fig~\ref{vlbamap}). The L-band flux density
measured in our VLBA image is consistent with that
measured by the FIRST survey. Therefore,  the core fraction is,
$c_{\rm f} \sim 1$, and the size of the VLBA beam is less than 30~pc at 
the redshift of the absorber.
The spin-temperature, measured from the ratio of 21-cm optical depth and
the $N$(H~{\sc i}) column density derived from the Lyman-$\alpha$ trough,
is 600$^{+220}_{-160}$~K which is consistent with the upper limit on $T_{\rm S}$ 
obtained from the width of the single component Gaussian fit to the 
21-cm absorption \citep{Srianand10}. 

\begin{table*}
\addtocounter{table}{1}
\caption{Summary of 21-cm searches in $z\ge2$ DLAs. Column 1: QSO. 
Column 2: absorption redshitft.
Column 3: log $N$(H~{\sc  i}). Column 4: Integrated optical depth. 
Column 5: Reference for  $\int\tau dv$
given in column 4. Column 6: the core fraction $c_f$. Column 9: the fraction
of CNM (see the text for its definition). Column 10: the H$_2$ fraction
and Column 11: References for $N$(H~{\sc i}) and/or f(H$_2$) measurements.
}
\begin{tabular}{lcccccccccc}
\hline
\multicolumn{1}{c}{QSO} & \zabs & log$N$(H~{\sc i}) &$\int \tau dv$ & Refs&$c_f$ & $T_{\rm S}/f_{\rm  c}$ & $T_{\rm S}$     & $f({\rm CNM})$& log f(\h2) & Refs\\
\multicolumn{3}{c}{} & (\kms) &\multicolumn{2}{c}{} & (K) & (K) \\ 
\multicolumn{1}{c}{(1)}        &  (2)  &      (3)       &       (4)           & (5) &  (6)   & (7) & (8) & (9)    &(10)&(11)  \\
\hline
\hline
\multicolumn{10}{c}{\bf 21-cm detections}\\
J020346.6+113445& 3.38714& 21.26$\pm$ 0.08 & 0.71$\pm$0.02 & 1         &0.76     &1397 &1062&$\sim$0.19 &$-6.2,-4.6$&a\\  
J031443.6+431405& 2.28977& 20.30$\pm$ 0.11 & 0.82$\pm$0.09 & 2         &....     &133  &....&$\sim$1.00 &....       &d\\
J044017.2$-$433309& 2.34747& 20.78$\pm$ 0.10 & 0.22$\pm$0.03 & 3         &0.59     &1493 & 881&$\sim$0.23 &....       &e\\  
J050112.8$-$015914& 2.03955& 21.70$\pm$ 0.10 & 7.02$\pm$0.16 & 4         &....     &390  & ...&....       &$\le -6.40$ &b\\  
J133724.6+315254& 3.17447& 21.36$\pm$ 0.10 & 2.08$\pm$0.17 & 5         &1.00     &600& 600&$\sim$0.33&$-7.00$      &c\\  
\multicolumn{10}{c}{\bf 21-cm non-detections having metallicity measurements}  \\
J033755.7$-$120412& 3.1799 & 20.65$\pm$ 0.10&$\le$0.06       & 6         & 0.62&$\ge$4057 &$\ge$2515 &$\le 0.08$ &$\le -5.10$  &b\\  
J033901.0$-$013318& 3.0619 & 21.10$\pm$ 0.10&$\le$0.06       & 6         & 0.68&$\ge$11435&$\ge$7775 &$\le 0.03$ &$\le -6.90$  &b\\  
J040733.9$-$330346& 2.569  & 20.60$\pm$ 0.10&$\le$0.12       & 7         & 0.44&$\ge$1807 &$\ge$795  &$\le 0.25$ & ....& e\\  
J040718.0$-$441013& 2.59475& 21.05$\pm$ 0.10&$\le$1.61       & 8         & ....&$\ge$380  &....      &     ....  &$-2.61^{+0.17}_{-0.20}$&b\\  
J040718.0$-$441013& 2.62140& 20.45$\pm$ 0.10&$\le$1.93       & 8         & ....&$\ge$80   &....      &   ....    &$\le -6.20$&b\\  
J043404.3$-$435550& 2.30197& 20.95$\pm$ 0.10&$\le$0.33      & 7         & ....&$\ge$1471 &....      &   ....    &$\le -5.15$&b\\  
J053007.9$-$250330& 2.81115& 21.35$\pm$ 0.07&$\le$0.58      & 9$^\dag$  & 0.94&$\ge$2103 &$\ge$1977&$\le 0.10$  &$-2.83^{+0.18}_{-0.19}$&b\\  
J091551.7+000713& 2.7434 & 20.74$\pm$ 0.10&$\le$0.37      & 7         & ....&$\ge$809  &....      &  ....     &.... &e\\  
J135646.8$-$110129& 2.96680& 20.80$\pm$ 0.10&$\le$0.33      & 6         & ....&$\ge$1042  &....      &  ....     &$\le -6.75$&b\\  
J135706.1$-$174402& 2.77990& 20.30$\pm$ 0.15&$\le$0.14      & 7         & ....&$\ge$777  &....      &  ....     &$\le -5.99$ &a\\  
J142107.7$-$064356& 3.44828& 20.50$\pm$ 0.10&$\le$0.14      & 7         & 0.69&$\ge$1231 &$\ge$849  &$\le 0.24$ &$\le -5.69$&b\\  
J234451.2+343348& 2.90910& 21.11$\pm$ 0.10&$\le$0.21      & 6         & 0.71&$\ge$3343 &$\ge$2373 &$\le 0.08$ &$\le -6.19$ &a\\  
\multicolumn{10}{c}{\bf 21-cm non-detections having no metallicity measurements}  \\                                          
J053954.3$-$283956& 2.9742 & 20.30$\pm$ 0.11&$\le$0.06      & 6         & 0.47 & $\ge$1812&$\ge851$ &$\le 0.23$ & ....&e\\  
J081618.9+482328& 3.4358 & 20.80$\pm$ 0.20&$\le$1.43      & 9         & 0.60 & $\ge$240 &$\ge144$ &$\le 1.00$ & ....&a\\  
J073320.4+272103& 2.7263 & 20.25$\pm$ 0.20&$\le$0.14      & 8         & .... & $\ge$692 &....     &.... & ....&a\\   
J080137.6+472528& 3.2235 & 20.80$\pm$ 0.15&$\le$0.22      & 8         & .... & $\ge$1563&....     &.... & ....&a\\  
J085257.1+243103& 2.7902 & 20.70$\pm$ 0.20&$\le$0.32      & 8         & 0.49 & $\ge$854 &$\ge418$ &$\le 0.48$ & ....&a\\
J101725.8+611627& 2.7681 & 20.60$\pm$ 0.15&$\le$0.29      & 8         & 0.81 & $\ge$748 &$\ge606$ &$\le 0.33$ & ....&a\\  
J124209.8+372005& 3.4135 & 20.50$\pm$ 0.30&$\le$0.11      & 8         & 1.00  & $\ge$1566&$\ge1566$&$\le 0.13$ & ....&a\\
J140501.1+041536& 2.708  & 21.07$\pm$ 0.24&$\le$0.19      & 9         & .... & $\ge$3369& ....    &.... & ....&f\\  
J140501.1+041536& 2.485  & 20.20$\pm$ 0.20&$\le$0.08      & 9         & .... & $\ge$1080 & ....    &.... & ....&f\\  
J140653.8+343337& 2.4989 & 20.20$\pm$ 0.20&$\le$0.31      & 8         & 0.76 & $\ge$279 &$\ge212$ &$\le 0.94$ & ....&a\\  
J143533.7+543559& 3.3032 & 20.30$\pm$ 0.20&$\le$0.26      & 8         & .... & $\ge$418 & ....    &.... & ....&a\\  
\hline
\end{tabular}
\begin{flushleft}
{
References in column \#5: 1) \citet{Kanekar07}, 2) \citet{York07}, 3) \citet{Kanekar06}, 4) \citet{Briggs89}, 5) \citet{Srianand10}, 
6) \citet{Kanekar03}, 7) \citet{Kanekar09ts}, 8) This paper, and 9) \citet{Curran10}. $^\dag$ Archival data from GBT08A\_003 (PI: Curran) was processed 
through our pipeline. See text for details. 
\\}
{References for $N$(H~{\sc i}) and/or f(H$_2$) measurements (column \# 11): a) This paper, b) \citet{Noterdaeme08}, c) \citet{Srianand10}, d) \citet{Ellison08}, e) \citet{Akerman05}
f) \citet{Curran10}.}
\end{flushleft}
\label{tablesum}
\end{table*}                                      

In Table~\ref{tablesum} we provide various details of our measurements 
together with the previous measurements at $z\ge 2$ from the literature. 
We present the results dividing the sample into three groups. These are systems
 with 21-cm detections (five systems), systems with 21-cm absorption upper limits with (twelve systems) and without (eleven systems) high-resolution optical 
spectra from which to derive accurate metallicities. The first
two groups are used to investigate the connection between UV measurements
and 21-cm optical depth. 
In all cases the 3$\sigma$ upper limits on the integrated
21-cm optical depth are computed assuming a line width of 10~\kms. 

The 21-cm detection rate from our sample, 
without putting any sensitivity limit, 
is 10\%. This is 13\% when we restrict to $\int \tau dv$ limit
of 0.4 \kms ( the limit achieved in the case of J1337+3152
where we have 21-cm detection).  Taken at face value, 
the extended sample listed in Table~\ref{tablesum}
gives a 21\% detection rate for  $\int \tau dv$ limit of 0.4 \kms.
For a $\int \tau dv$ limit of 0.2 \kms we get the detection rate of
28\%.
However, these may not be representative values as the list of 
systems compiled from the literature may be biased towards detections 
as some authors may not have reported their non-detections systematically

Since we know $N$(H~{\sc i}) from the damped Lyman-$\alpha$ line, the 
detection limit on the integrated optical depth implies a lower limit on the 
ratio $T_{\rm S}/f_{\rm c}$. 
The  $T_{\rm S}/f_{\rm c}$ measurements are reported in column 7 
of Table~\ref{tablesum}. 
In column 6 of this table, we give the core fraction $c_{ f}$. 
As mentioned above, $c_f$ is basically the ratio of flux density in the unresolved core seen in
VLBA images to the total flux density measured in the arcsecond scale FIRST images. 
For the objects from the literature we use the $c_f$ values given in
\citet{Kanekar09vlba}. These measurements were made at 327~MHz, close
to the redshifted 21-cm frequencies. 
Following \citet{Kanekar09vlba} we use core fraction ($c_f$)
as the estimate of the covering factor ($f_c$).
The $T_{\rm S}$ measurements given in column 8 of Table~\ref{tablesum}
are obtained by assuming $f_c = c_f$.

In Fig.~\ref{fwnm} we plot the  percentage of DLAs having $T_{\rm S}$ 
greater than a limiting value $T_{\rm S}^l$ as a function of 
$T_{\rm S}^l$ for systems with $T_{\rm S}$ measurements given in column
8 of Table~\ref{tablesum}. The lower envelope of the shaded region is 
obtained considering all the lower limits on $T_{\rm S}$  as 
measurements. The upper envelope is obtained assuming
all the lower limits as measurements with $T_{\rm S}\ge T_{\rm S}^l$.
It is clear from the figure that
more than 50\% of the DLAs have  $T_{\rm S}\ge 700 K$. Remember that the $T_{\rm S}$ measured in an individual DLA
is the harmonic mean temperature of different phases that contribute to the observed
$N$(H~{\sc i}).
Assuming that the gas is simply a two phase medium with similar 
covering factors the
fraction of H~{\sc i} in the CNM (called $f{\rm (CNM)}$) can be written as,
\begin{equation}
f{\rm (CNM)} = {1 \over T_{\rm S}^W} \bigg{[} {T_{\rm S}^{\rm C}T_{\rm S}^W \over T_{\rm S}} - T_{\rm S}^{\rm C}\bigg{]} 
\end{equation}
where,  ${T_{\rm S}^{\rm C}}$ and ${T_{\rm S}^{\rm W}}$ are the spin-temperature of the CNM and WNM
respectively.
\citet{Srianand05} have noticed that the H~{\sc i} phase traced by the H$_2$ 
absorption has temperature typically in the range 100-200~K. Thus we
consider the CNM temperature to be 200 K (instead of 70 K as
seen in CNM of the
Galaxy) so that the $f$(CNM) we get will be
a conservative upper limit.  Assuming ${T_{\rm S}^{\rm C}}\sim200$~K and  ${T_{\rm S}^{\rm W}}\sim10^4$~K,
$T_S = 700$ K can be obtained for a combination of $f{\rm (CNM)}$ = 0.27 and   
$f{\rm (WNM)}$ = 0.73. Therefore  $f{\rm (CNM)}$ is less than  0.27 in 
at least 50\% of the DLAs. Note that choosing  ${T_{\rm S}^{\rm W}}\sim8000$~K
(as suggested for the Galactic ISM) instead of the 10$^4$ K used here, 
does not change the results appreciably. 

We estimate $f{\rm (CNM)}$ for the four 21-cm detections
(excluding J0501-0159 (B0458-020)  for which we do not have the covering factor value). 
Apart from J0314+4314 (3C082) which seems to be a special case \citep{York07}, 
the CNM seems to represent roughly 20 to 
30\% of the total $N$(H~{\sc i}) measured in these DLAs. 
For individual non-detections, we can calculate conservative upper limits of the 
fraction of $N$(H~{\sc i}) in the CNM phase assuming $T_{\rm S}^{\rm C}$~=~200~K.
The values of $f$(CNM) are given in column \#9 of Table~\ref{tablesum} for systems with 
$f_{\rm c}$ measurements. The upper limits vary between 0.10 and 1.0 with a
median value of 0.23. 
Thus the analysis presented here, under the assumption that $f_c = c_f$, 
suggests that most of the neutral hydrogen in 
high-$z$ DLAs is warm. This is very much consistent with the conclusion of \citet{Petitjean00} 
based on the lack of \h2 detections in most high-$z$ DLAs.
\begin{figure}
\centerline{
\includegraphics[scale=0.4,angle=0.0]{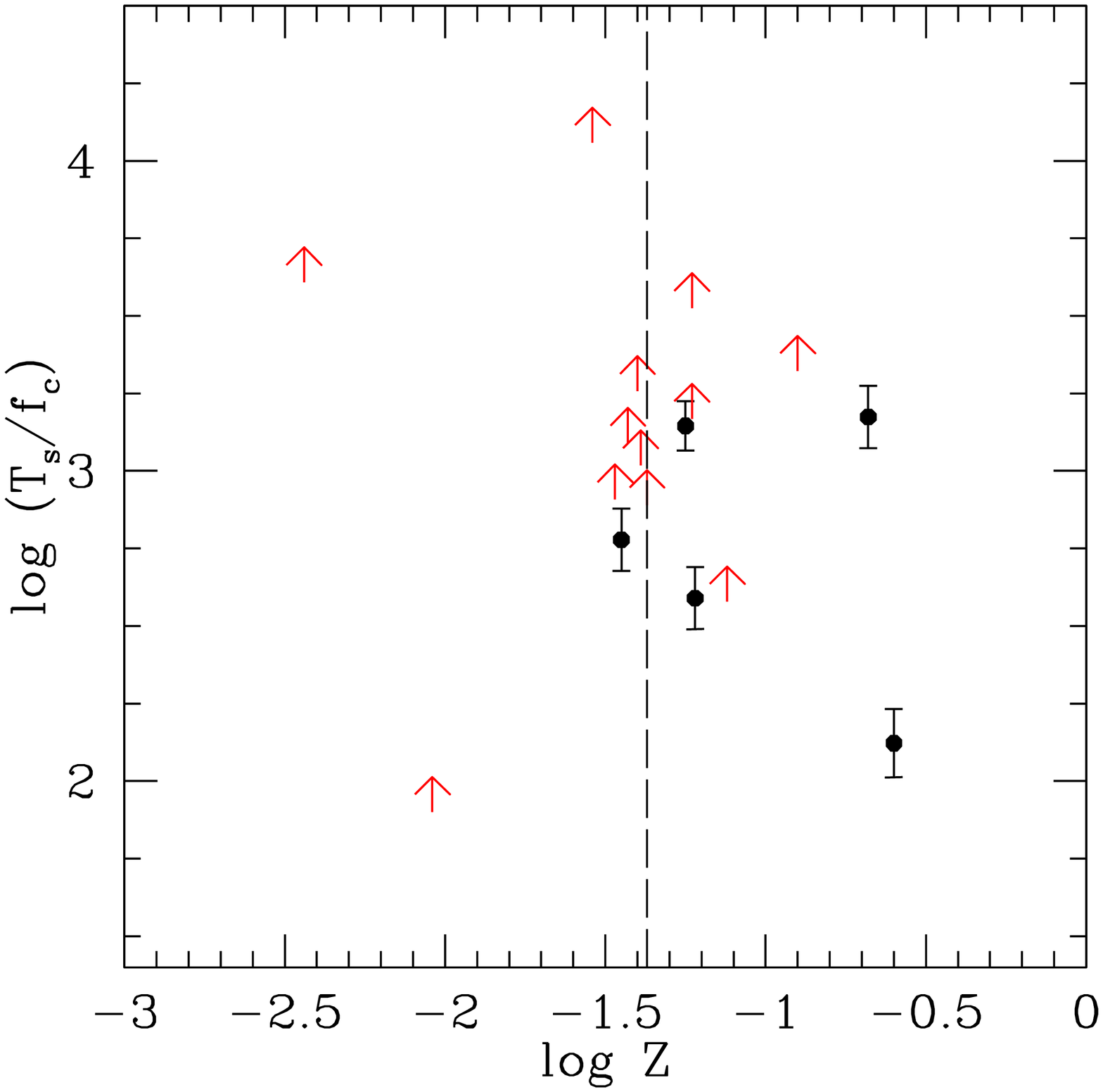}
}
\caption{Metallicity vs $T_S/f_c$. The vertical dashed line marks the
median metallicity measured in our sample. }
\label{tsvmet}
\end{figure}

\section{Results of Correlation analysis}
\label{gencor}
\begin{figure*}
\centerline{
\vbox{
\hbox{
\includegraphics[scale=0.4,angle=0.0]{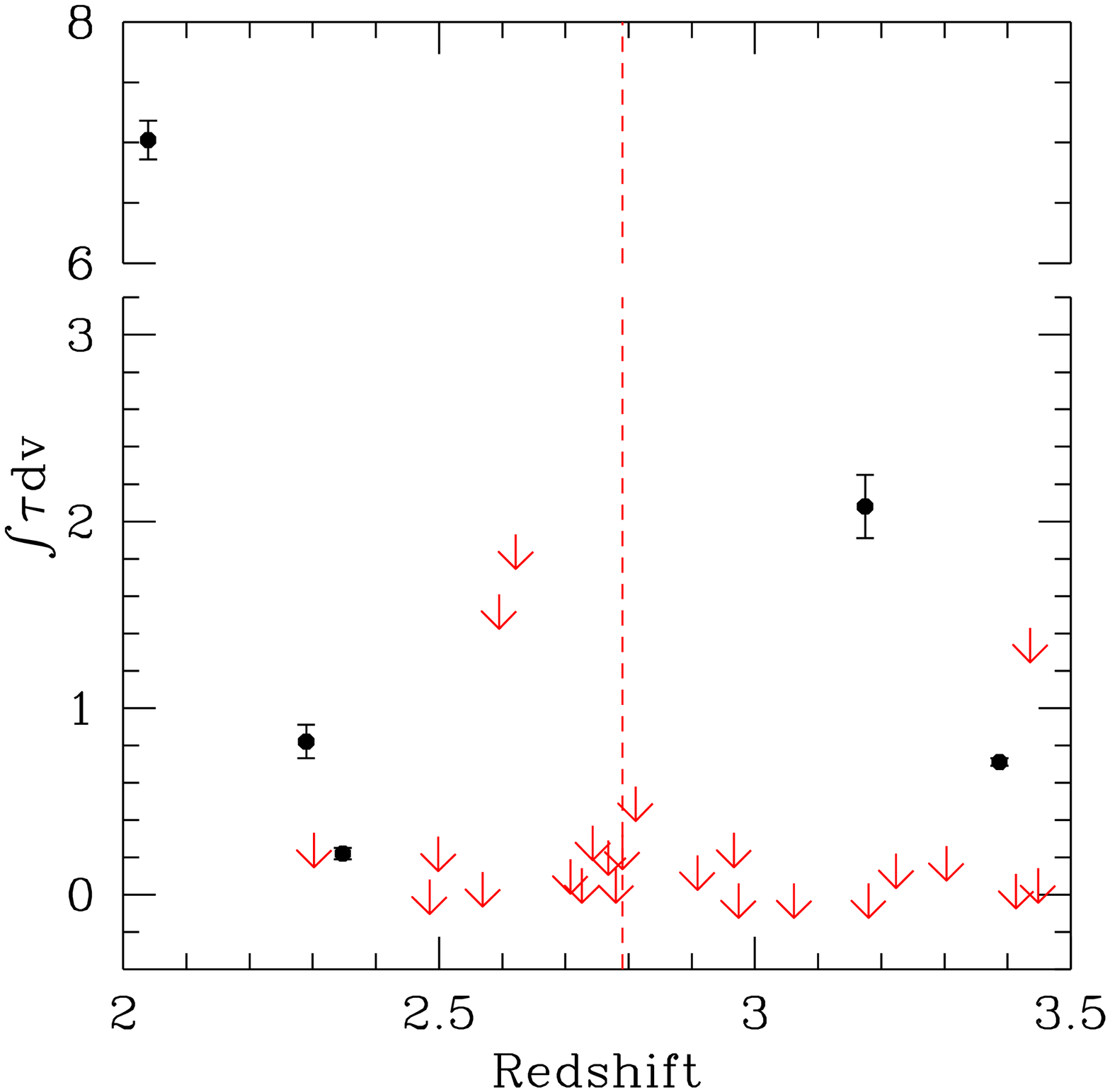}
\includegraphics[scale=0.4,angle=0.0]{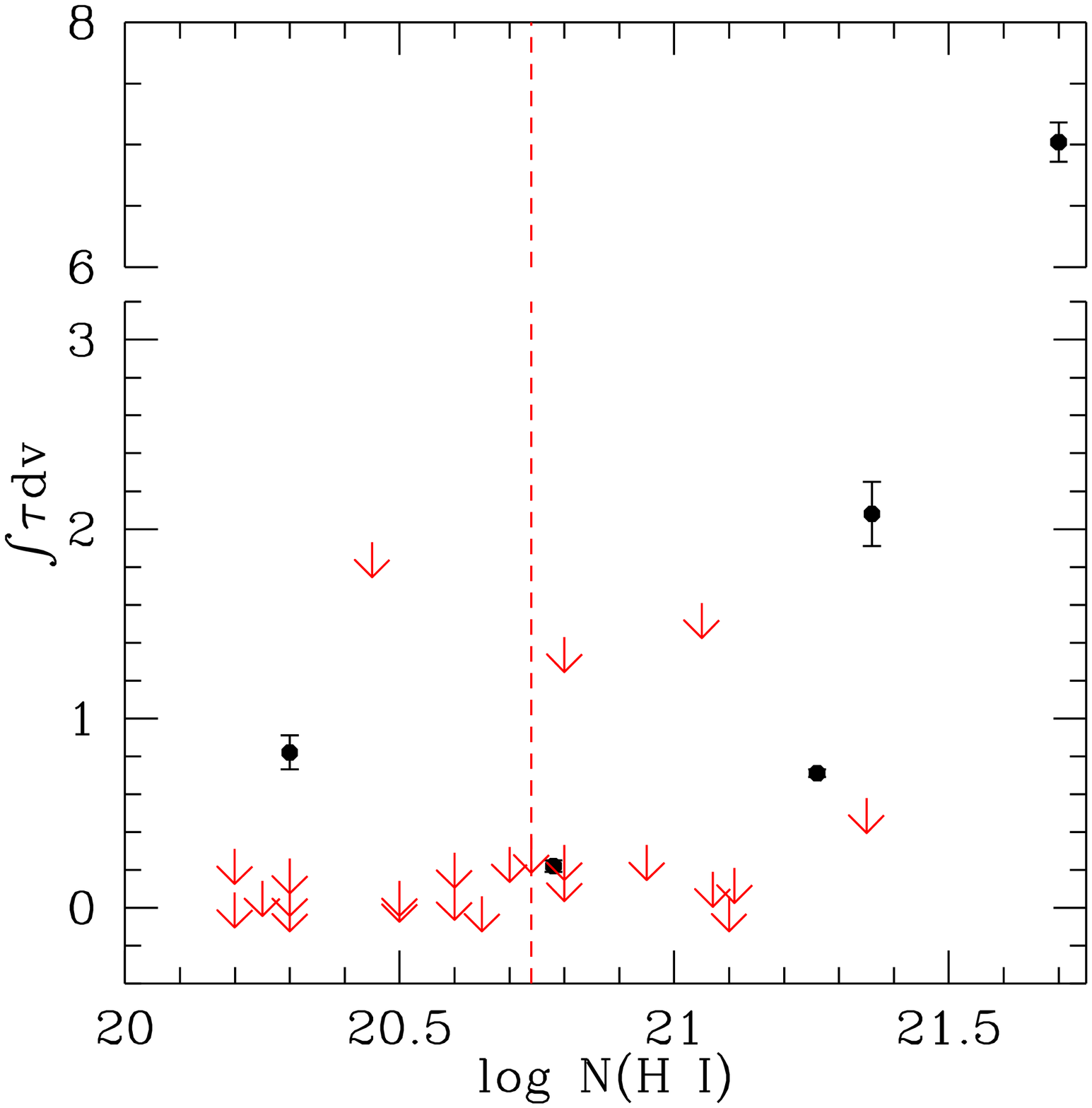}}
\hbox{
\includegraphics[scale=0.4,angle=0.0]{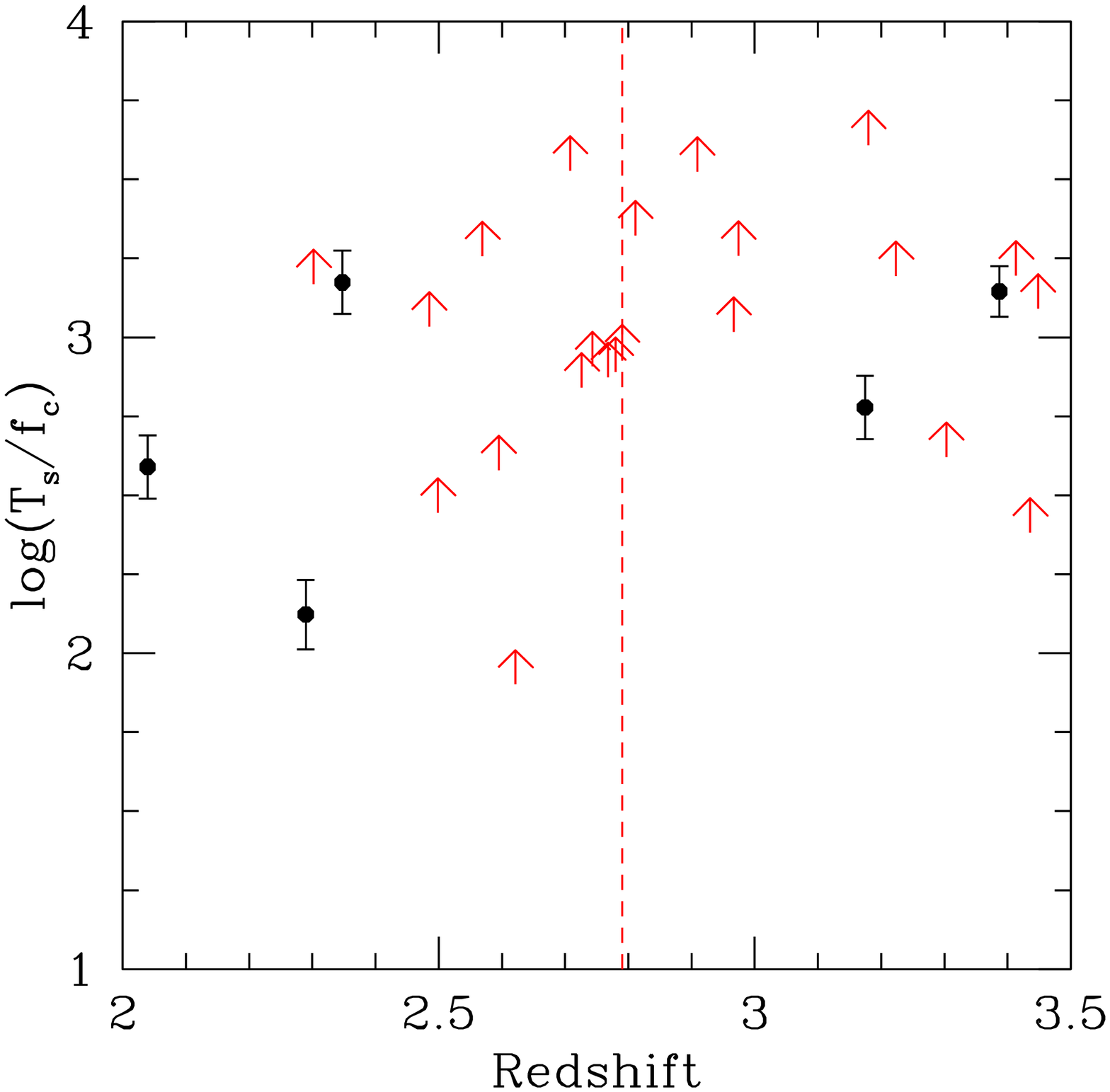}
\includegraphics[scale=0.4,angle=0.0]{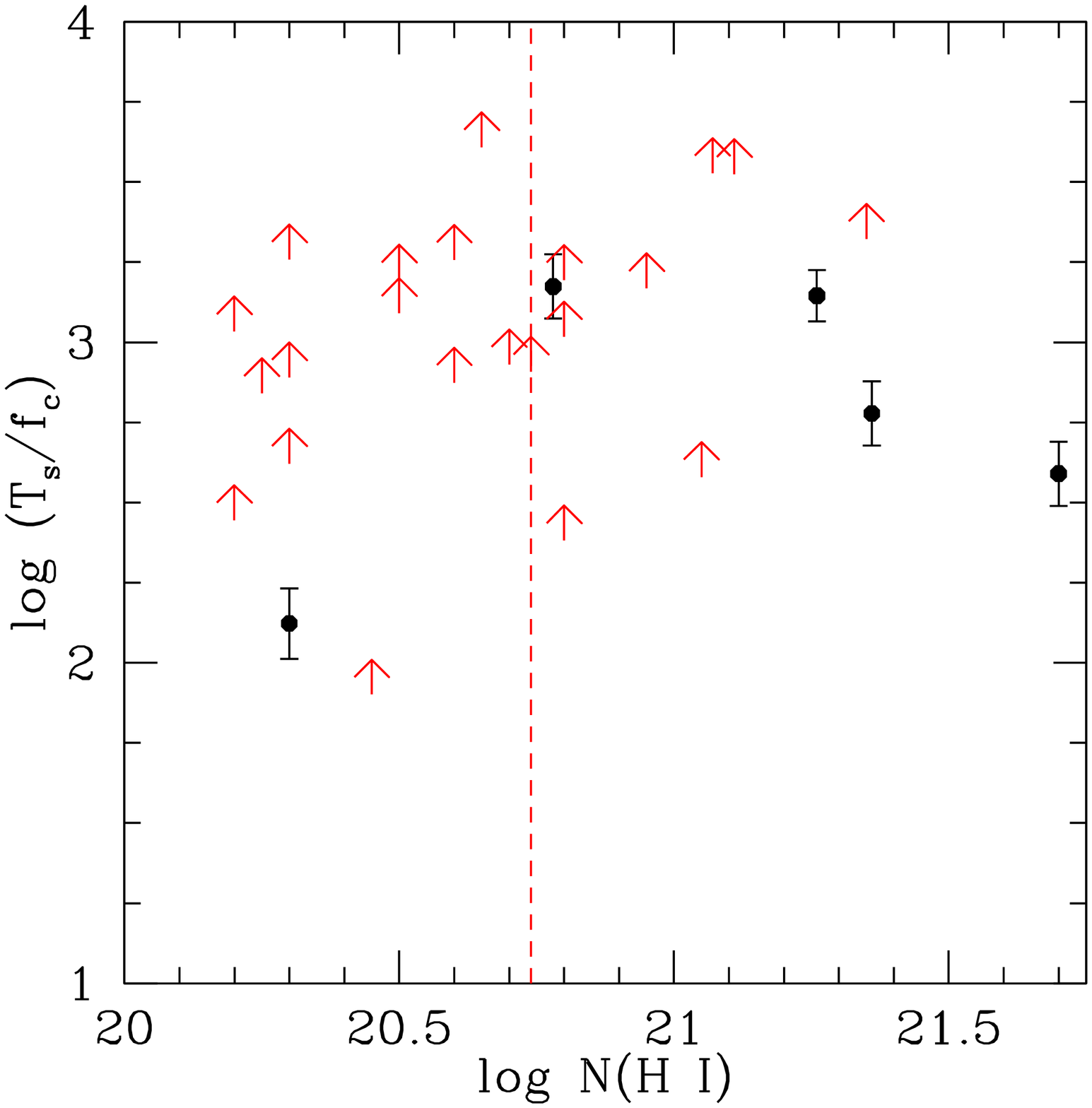}
}
}}
\caption{Properties of 21-cm absorptions vs redshift and $N$(H~{\sc i}).
The vertical dashed lines give the median value of the quantity plotted
in the x-axis.}
\label{figcor}
\end{figure*}

In this Section we explore correlations between the 21-cm
optical depth and other observable parameters. As we have only
a few 21-cm detections and mostly 
upper limits we use survival analysis and in particular the generalised rank 
correlation test \citep{Isobe86}. For this purpose we use
the Astronomical SURVival analysis (ASURV) package.

\subsection{Metallicity vs T$_s$/$f_c$}
Firstly we study the importance of the metallicity of the gas.
Only 3 DLAs in our sample have measurements of metallicity from
high resolution optical spectroscopy. In the extended sample
at $z\ge2$ (see Table~\ref{tablesum}) there are 17 systems with metallicity measurements
and 21-cm spectra. In Fig.~\ref{tsvmet}, we plot $T_{\rm S}/f_{\rm c}$ versus
metallicity.  The vertical long-dashed line marks the median metallicity of the points 
plotted in the figure. The only detection found in the low metallicity half is for 
\zabs~=~3.1745 towards J1337+3152 reported from
our survey.  The other four detections are from the high metallicity half. 
The non-parametric generalized Kendall rank correlation test 
suggests only a weak correlation between Z and $T_{\rm S}/f_{\rm c}$ (at the 1.42$\sigma$ level) with the probability that 
it can arise due to chance being 0.15. The significance is even lower (i.e 0.9$\sigma$ with
a chance probability of 0.37) when we use $T_{\rm S}$ (instead
of $T_{\rm S}/f_{\rm c}$) for cases where we have estimated 
$f_{\rm c}$ measurements. 
We wish to point out that a correlation between $T_s/f_c$ and metallicity
is reported in the literature \citep[][]{Curran07,Kanekar09ts,Curran10}.
The lack of correlation in our sample (with systems in a restricted
redshift range) could either reflect redshift evolution of the relationship
or small range in metallicity covered by the sample.
Metallicity measurements for the remaining 11 systems in Table~\ref{tablesum} would 
allow us to address this issue in a statistically significant manner.

We also looked at the possible correlation between the 21-cm
optical depth and the velocity width ($\Delta v$) of the low 
ionization lines. This information is available for 14 sources.
Again we find no statistically significant correlation 
between the two.  This is inconsistent with the 2.2 to 2.8 $\sigma$ level
weak correlation between $W$(Mg~{\sc ii}) and $\Delta v$
reported by \citet{Curran07}.
However, this is consistent with the finding of
\citet{Gupta09}, that
$W$(Mg~{\sc ii}) 
and 21-cm optical depth are not correlated for a sample of 33 strong Mg~{\sc ii} systems at 
1.10$\le z \le 1.45$ \citep[see also][]{Kanekar09mg2}. 

\subsection{Redshift dependence}

In the left hand side panels of Fig.~\ref{figcor} we plot 
$T_s/f_c$ and integrated 21-cm optical depth  vs redshift.
No clear correlation is evident in this figure. The non-parametric
Kendall test finds no significant correlation between $\int \tau dv$ 
(or T$_S$) and $z$. Note our sample probes only  a restricted 
redshift range in terms of time interval probed.
However, the lack of correlation found here is consistent
with the near constancy of T$_s$/$f_c$ as a function of redshift
found by \citet{Curran10}. Understanding the redshift dependence of
$T_S$ is very important in particular to address whether there is 
any evolution in $T_s$ \citep[][]{Kanekar03} or
geometric effects \citep[][]{Curran06}. 
To make an unbiased comparison we need to have 21-cm measurements 
at low $z$ for a well defined sample of DLAs detected 
based on \lya\ absorption.

\subsection{Dependence on $N$(H~{\sc i})}

Recently \citet{Curran10} have found a 3$\sigma$ level
correlation between $N$(H~{\sc i}) and $T_s/f_c$. To check whether
this correlation holds at $z>2$, we plot, in the top  panels 
of Fig.~\ref{figcor}, the integrated
21-cm optical depth as a function of redshift and $N$(H~{\sc i}).
We note
that there is a tendency for more 21-cm detections in DLAs
with higher $N$(H~{\sc i}). 
However, the non-parametric
Kendall test finds no significant correlation  between
$\int \tau dv$ and  $N$(H~{\sc i}).
In the bottom right panel we plot $T_{\rm S}/f_{\rm c}$ against log~$N$(H~{\sc i}). 
The Kendall test does not show any significant relation between the two 
quantities (1.28$\sigma$ with a probability of 0.2 for this to be due to chance). 
Thus we do not find any evidence for the 21-cm optical depth to depend on
$N$(H~{\sc i}) in our sample.

\section{21-cm absorption and H$_2$}
\label{mole}
As 21-cm absorption and H$_2$ molecules can give complementary information
on the physical state of the gas. In this Section, we study the relationship
between these two indicators. 
There are 13 DLAs in our extended sample for which
the expected optical wavelength range of redshifted \h2 
absorptions has been observed at high spectral resolution. 
Nine of these sources are part of the UVES sample of \citet{Noterdaeme08}. 
\citet{Srianand10} have reported the detection of \h2 in J1337+3152 
and here we report the search 
for \h2 in the remaining three DLAs (\zabs = 3.3871 towards 
J0203+1134, \zabs = 2.7799 towards J1357$-$1744
and \zabs = 2.9091 towards J2344+3433). In the 10th column of 
Table~\ref{tablesum}, we summarize 
the molecular fraction 
$f$(\h2)~=~2$N$(\h2)/(2$N$(\h2)+$N$(H~{\sc i}))] derived for these
13 systems.

In 8 systems, neither 21-cm absorption nor H$_2$ molecules are detected
with typical upper limits of the order of 10$^{-6}$ for $f$(\h2).
Apart from the system at \zabs~=~2.6214 towards J0407$-$4410, the lower
limits on $T_{\rm S}/f_{\rm c}$ for the remaining 7 systems are higher 
than 700~K.  
There are 4 cases where $f_{\rm c}$ measurements are available. In three
cases (\zabs~=~3.1799 towards J0337$-$1204, \zabs~=~3.0619
towards J0339$-$0133 and \zabs~=~2.9019 towards J2344+3433), the 
lower limit on $T_{\rm S}$ is more than 2000~K. These are in line
with the suggestion by \citet{Petitjean00} that the absence of
\h2 in most of the DLAs is due to the low density and high
temperature of the gas.

In two cases (\zabs~=~2.5947 towards J0407$-$4410 (CTS 247) and \zabs~=~2.8112 
towards J0530$-$2503 (PKS 0528-250)), strong H$_2$ absorption is detected with 
rotational excitations consistent with the \h2-bearing gas being a CNM.
However, 21-cm absorption is not detected in either case. We
discuss these two systems in detail below.  

Among the five 21-cm absorbers, high resolution UVES spectra covering 
the expected wavelength range of \h2 absorption are available for four 
systems. The exception is the \zabs~=~2.28977 
system towards J0314+4314 (B0311+430). For the \zabs~=~2.3474 system towards 
J0440$-$4333 (B0438$-$436) the continuum flux in the expected wavelength range is removed 
by high ionization lines from an associated system, as well as by a high-$z$ Lyman 
limit system present along the line of sight.
Below we discuss the five systems where simultaneous
analysis of \h2 and 21-cm absorption is possible.

\begin{figure*}
\centerline{
\includegraphics[scale=0.6,angle=270]{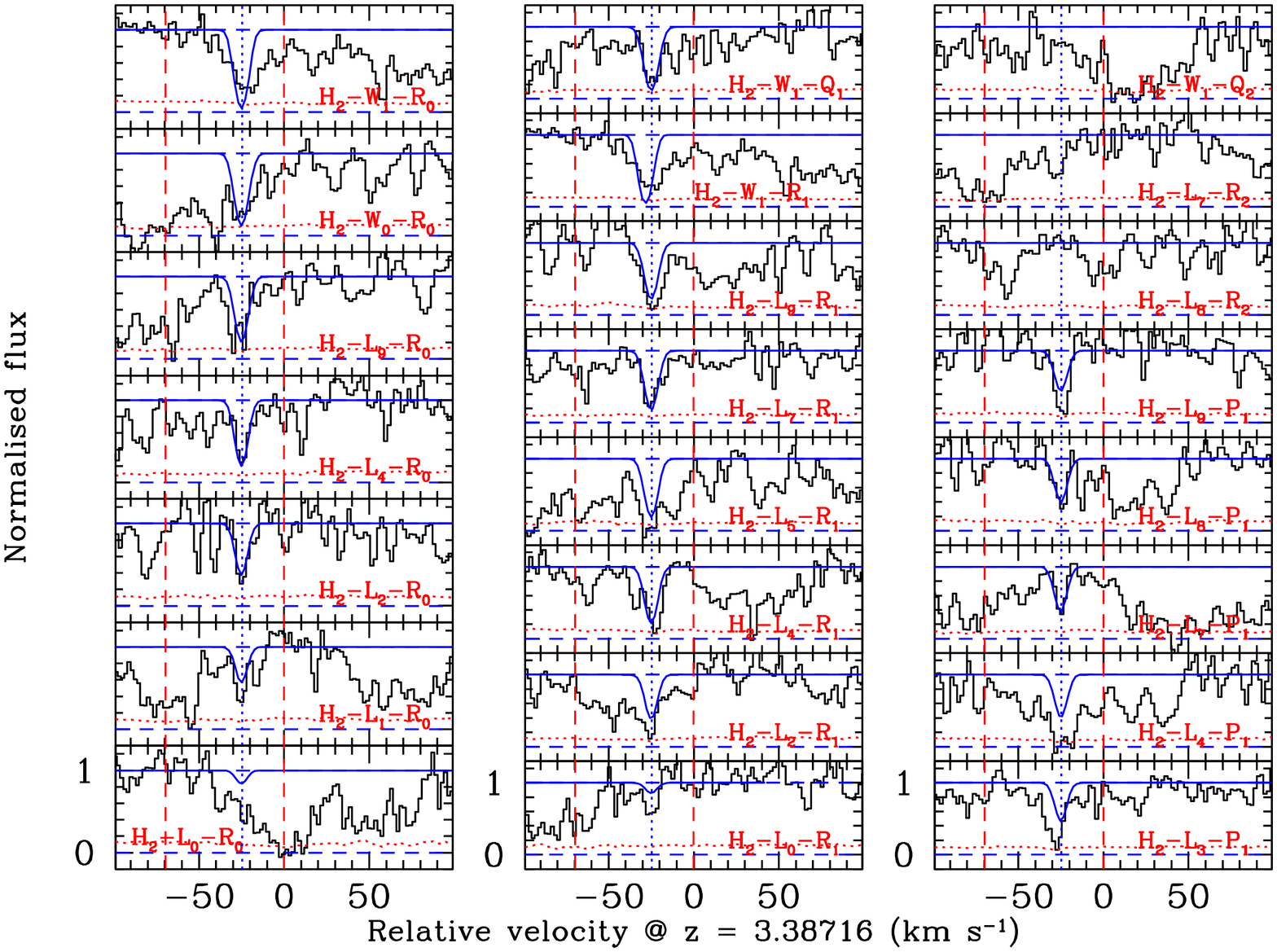}
}
\caption{Voigt profile fits to H$_2$ Lyman and Werner band
absorption lines in the \zabs~=~3.3868 DLA system towards J0203+1134 (PKS~0201+113).
The zero of the velocity scale is defined at $z = 3.38716$. The two
vertical lines at $v = 0$ and $-$68~\kms~ show the locations of
two 21-cm absorption components reported by \citep{Kanekar07}. 
The vertical dotted line indicates the location of \h2 absorption.
 In each panel we also show the error spectrum with dotted curves.
}
\label{h2b0201}
\end{figure*}

\subsection{\zabs = 3.3868 DLA towards  J0203+1134 (PKS 0201+113)}

Searches for 21-cm absorption in this system  have yielded
conflicting results \citep{DeBruyn96, Briggs97}.
Based on GMRT spectra taken at three different epochs, 
\citet{Kanekar07} reported the detection of 21-cm absorption 
in two components at \zabs~=~3.387144(17) and 3.386141(45). 
Using $N$(\ion{H}{i})$\sim$ (1.8$\pm$0.3)$\times 10^{21}$~cm$^{-2}$ they 
obtained $T_{\rm S} = [955\pm160](f_{\rm c}/0.69)$~K. 
Using high resolution optical spectrum,
\citet{Ellison01Q0201} have
found a gas phase metallicity of 1/20 of solar with
very little dust depletion.
The gas cooling rate,
log~$l_{\rm c} = -26.67\pm0.10$~erg s$^{-1}$ Hz$^{-1}$,
derived using C~{\sc ii*} absorption is consistent with
this DLA being part of high-cool population defined by
 \citet{Wolfe03a}. 
From \citet{Kanekar07} we 
can see that the strongest 21-cm absorption does not correspond to the strongest velocity
component in  either C~{\sc ii$^*$} or Fe~{\sc ii}. 

Here we report the detection of  
H$_2$ absorption from J=0 and J=1 levels originating 
from both Lyman and Werner bands 
(see Fig.~\ref{h2b0201}). A single component Voigt profile fit reproduces the data well. As the 
Lyman-$\alpha$ forest is dense and the spectral signal-to-noise ratio is not very high due to the
faintness of the QSO, we considered a range of $b$ values (i.e between 1 and 5~\kms) to get the best
fit values of log[$N$(H$_2$, J=0)] in the range 16.10$-$14.48 and  log[$N$(H$_2$, J=1)] = 16.03$-$14.57. 
We estimated the kinetic temperature using the ortho-to-para ratio (i.e $T_{01}$) 
and found it to be in the range 48$-$108~K for the range of $b$ parameters
considered above.  We note that for $b$ parameters greater than 2~\kms, the H$_2$ lines are 
mainly in the linear portion of the curve of growth and the column density estimate is insensitive to the 
assumed value. 
The average molecular fraction, $log~f({\rm H_2})$, in the range, 
$-4.6\le log~f({\rm H_2})\le -6.2$. 

Despite the gas being cold, there is no 21-cm absorption
detected at the position of the H$_2$ component (at z = 3.38679) 
which is well separated from the 21-cm absorption components 
If we use $f_c=0.76$, as found by \citet{Kanekar09vlba} using 326 MHz
observations, we find
log~$N$(H~{\sc i}) $\le19.12$. This is less than 1\% of the total H~{\sc i} column density measured in this system.

Unlike most of the strong \h2 systems, this system does not  show detectable C~{\sc i} absorption. This means we do not have,
unfortunately, an independent estimate of the density 
from fine-structure excitation.

\subsection{\zabs = 2.5948 towards J0407-4410 (CTS 247)}

 As the radio source is faint, our GBT spectrum only gives 
a weak limit on the spin temperature, $T_{\rm S} \ge 380$~ K when we use a 
line width of 10 \kms.
\citet{Srianand05} have reported log~$N$(C~{\sc ii$^*$})~=~13.66$\pm$0.13. This, together with log~$N$(H~{\sc i})~=~21.05$\pm$0.10, 
gives a gas cooling rate of  log~$l_{\rm c} = 26.92\pm0.16$. This is very close to  the value $l_{\rm c}^{\rm crit}$
that seems to demarcate between the high and low cool systems defined by \citet{Wolfe08}. 

\citet{Ledoux03} reported the detection of H$_2$ from this system.
The H$_2$ absorption is well fitted with two components at \zabs~=~2.59471 and 2.49486
with log~$N$(H$_2$)~=~18.14 and 15.51 respectively \citep{Srianand05}. These components have $T_{01}$~=~121$\pm$10 
and 91$\pm$6~K respectively. The average molecular fraction, log~$f$(H$_2$), is found to be $-2.42^{+0.07}_{-0.12}$ with 
an average metallicity of $-1.02\pm0.12$ and moderate dust depletion \citep{Ledoux03}. 

The absence of 21-cm  absorption from this system is intriguing
as H$_2$ components have T$\sim$100 K.
With the same $b$ parameters as used to fit the \h2 lines and 
the rms from the GBT spectrum, we get a 3$\sigma$ upper limit
of $\int \tau dv = 0.88$ \kms.  This translates to a constraint, $f_{\rm c}\times$$N$(H~{\sc i}) $\le$ 2$\times 10^{20}$~cm$^{-2}$
in the H$_2$ components where we have assumed $T_{\rm S}$~=~$T_{01}$.
Unfortunately we do not have a VLBA image of this source
and it is difficult to constrain the covering factor of the gas. 
If we assume $f_{\rm c}\sim 1$ then the upper limit on $N$(H~{\sc i}) implies 
that the H$_2$ component is a sub-DLA with
log~$f$(\h2) $>-1.85$.

 From the column densities of the C~{\sc i} fine-structure lines,
\citet{Srianand05} have constrained the particle density in the gas to be in the range 
$4.5<n_{\rm H} ({\rm cm}^{-3})<57.3$. For $n_{\rm H} = 4.5~{\rm cm}^{-3}$ and $f_{\rm c} = 1$, we estimate 
the thickness of the \h2 cloud (i.e $N$(H~{\sc i})/$n_{\rm H}$ ) 
along the line of sight to be $\le 15$ pc.

\subsection{\zabs = 2.0395 towards J0501$-$0159 (B0458-020)}

21-cm absorption in two velocity components was reported by \citet{Wolfe85}. The background radio source shows structure 
over a wide range of scales and the absorbing cloud seems to cover most of these components \citep{Briggs89}. 
The estimated extent of the H~{\sc i} absorber is $\sim$8~kpc and the spin-temperature of the system is 390~K.

C~{\sc ii$^*$} absorption is detected  and the measured cooling rate is $l_c = -26.41\pm0.10$ \citep{Wolfe08}. This is consistent with the high cool population. 
\h2 molecules are not detected with an upper limit of the molecular fraction of log~$f$(H$_2$)~$\le -6.52$. This is one
of the rare DLAs to show Lyman-$\alpha$ emission in the middle of the DLA absorption \citep{Moller04,Heinmuller06}.

Based on the star formation rate derived from the Lyman-$\alpha$ flux, \citet{Heinmuller06} argued that the ambient radiation field
is 10 times higher than the Galactic UV background. This excess radiation could be the main reason for the absence of \h2 in the
gas. 
This picture is also confirmed by the lack of associated C~{\sc i} absorption.

\subsection{\zabs = 2.8111 towards  J0530-2503 (PKS 0528$-$250)}

\citet{Carilli96} searched for 21-cm absorption in this system using the WSRT and did not find any significant absorption. 
Recently \citet{Curran10} have reported  GBT observations of this source with a better 
signal-to-noise. They obtain a lower limit of the spin temperature of 700~K for $f_{\rm c} = 1$, 
again assuming a width of 10 \kms. This source was found to be compact in the low frequency VLBA images with $f_c = 0.94$ \citep{Kanekar09vlba}.

No measurement of cooling rate is possible for this system as the 
the C~{\sc ii$^*$} absorption is blended with other saturated lines.

This system has log~$N$(H~{\sc i})~=~21.35$\pm$0.07, $Z$~=~$-0.91\pm0.07$ and \zabs~higher than \zem.
This is the first high redshift DLA where \h2 absorption has been detected \citep{Levshakov85,Srianand98, Srianand05}.
The high resolution UVES spectrum reveals two strong \h2 components at \zabs~=~2.81100 and 
2.81112 with log~$N$(\h2) ~= ~17.93$^{+0.14}_{-0.20}$ and 17.90$^{+0.11}_{-0.14}$, and 
$ T_{01}$~=~167$\pm$7 and 138$\pm$12 K respectively.
%


We re-reduced the GBT data of \citet{Curran10} in the same way as our GBT data and found the rms in the frequency range 
expected for the 21-cm absorption associated with the \h2 components to be 6.7~mJy.  This, together with the temperature 
$T_{01}$ and the \h2 $b$ parameter give a constraint of log~$N$(H~{\sc i})~$\le$~20 in the \h2 components. 

C~{\sc i} absorption is detected only 
in one of the \h2 components (i.e at \zabs~=~2.81112)\citep{Srianand05}.  From the C~{\sc i}  level populations 
and for the above temperature, we derive that the particle density is in the range $25<n_{\rm H} ({\rm cm}^{-3})<270$. 
If we use the $N$(H~{\sc i}) limit of 10$^{20}$~cm$^{-2}$ derived from the 21-cm absorption, and use the constraint we get 
for the hydrogen density in the case of the \zabs = 2.81112 component, we get a limit on the thickness of the gas along
the line of sight to be 1.33~pc. 

\subsection{\zabs= 3.1745 towards J1337+3152}

Detection of 21-cm and \h2 in this system was reported in \citet{Srianand10}. The weak \h2 absorption is  
well aligned with the strongest metal line component but is
slightly shifted by 2.5 \kms with respect to the 21-cm absorption component.
This again suggests that the two transitions do not originate exactly from the same gas and that 
the absorptions arise from an inhomogeneous absorbing region.

As the \h2 column density is very low and the ratio of $N$(\h2,J=0) to $N$(\h2,J=1) is close to the maximum value
allowed by the Boltzmann distribution,  we cannot constrain the kinetic temperature of the gas.
C~{\sc ii$^*$} absorption is detected with a total column density log~$N$(C~{\sc ii}$^*$)~=~13.61$\pm$0.08.
Using log~$N$(\hi)~=~21.36$\pm$0.10 we derive a gas cooling rate of log $l_{\rm c} = -27.28\pm0.13$. Thus the system 
belongs to the low cool part of the DLA population.
The quasar is unresolved in our 1420 MHz VLBA image with a total flux density consistent with the measurement from FIRST
observations. This suggests that most of the emission is in the unresolved component. Using Gaussian fits we
estimate the limit on the largest angular size to be $\le$ 3.8 mas  (or $\le$~30 pc at \zabs).
Given the compact nature of the radio source it is  intriguing to see the difference between the 
\h2 and 21-cm absorptions.  
This clearly shows that the absorbing gas contains a mixture of different phases at parsec scales.

\section{Results and discussion}
\label{results}

We have carried out a systematic search for 21-cm absorption in 10 
DLAs at \zabs$>2$ using GMRT and GBT. 
We detect 21-cm absorption in only one of them. From our sample we 
find the 21-cm detection rate is 13\% 
for a $\int \tau dv$ limit of 0.4 km/s (the detection limit reached in 
the case of J1337+3152).
We also obtained 1420~MHz VLBI images for the  sources in our sample.

The 21-cm detection at $z\ge 2$ seems to favour systems with high 
metallicity  and/or  high $N$(H~{\sc i}) \citep[see also][]{Kanekar09ts, Curran10}.
This basically means that the probability of detecting cold components that can 
produce detectable 21-cm absorption is higher in systems with high values 
of $N$(H~{\sc i}) and Z.  However, we do not
find any correlation between the integrated optical depth 
(or T$_{\rm S}$/$f_{\rm c}$) and $N$(H~{\sc i}) or metallicity.

It is important to address the covering factor issue before drawing
any conclusions on $T_S$. Ideally one should do high spatial resolution
VLBA spectroscopy for this purpose \citep[see for example][]{Lane00}.
However, this is not possible at present specially for $z\ge2$ absorbers.
Therefore, we proceed by  assuming that the core fraction  found in the VLBA 
images
as the covering factor of the absorbing gas \citep[as in the case of][]{Kanekar09vlba}.
We find that more than 50\% of DLAs have weighted mean spin  temperature 
($T_{\rm S}$) in excess of 700 K. For the  assumed temperature
of the CNM gas $T_{\rm S}^C = 200$ K (as seen in \h2 components in high-z DLAs) 
we find that more than 73\% of H~{\sc i} in such systems is
originating from WNM. The median value CNM fraction (i.e $f$(CNM)) 
obtained for the detections and the
median value of upper limits in the case of non-detections are in the range 0.2 to 0.25. 

We study the connection between 21-cm and \h2 absorption in a sub-sample of 13 DLAs where both these species can be searched for.
We report the detection and detailed analysis of \h2 molecules in the \zabs=3.3871 DLA system towards J0203+1134 where 21-cm 
absorption is also detected. For a $b$ parameter in the range 1-5~\kms\  we find 14.57$\le$log~$N$(\h2)$\le$16.03. The inferred kinetic
temperature is in the range 48-108~K based on $T_{01}$ of H$_2$. However no 21-cm absorption is detected at the very position of 
this \h2\  component. This suggests that the H~{\sc i} column density associated with this component is $\le$ 10$^{19}$~cm$^{-2}$. 
However, the lack of proper coincidence between 21-cm and any of the strong 
UV absorption components may also mean that 
the radio and optical sight lines probe different volumes of the gas.

In the case of 8 DLAs, neither 21-cm nor H$_2$ are detected. Typical upper 
limits on the molecular fraction  ($f_{\rm H_2}$) in these systems are 
$\le 10^{-6}$. 
The lack of \h2 in DLAs can be explained if the H~{\sc i} gas originates
from low density regions photoionized by the metagalactic UV
\citep[see for example,][]{Petitjean92,Petitjean00,Hirashita05}.
This also indicates that the volume filling  factor of \h2 in DLAs 
is small  \citep{Zwaan06}.  Typical limits obtained for  $T_{\rm S}$
in these systems are consistent with only a small fraction of
the H~{\sc i} gas originating from the CNM phase as suggested by the
lack of \h2 absorption.

In two cases strong \h2 absorption is detected and kinetic temperatures are in the range 100-200~K, but 21-cm absorption is not 
detected.  Even in two cases where both the species are  detected they do not originate from the same velocity component. 
The lack of 21-cm absorption directly associated with \h2\  indicates that only a small fraction (typically $\le$ 10\%)
of the neutral hydrogen seen in the DLA 
is associated with the \h2 components \citep[see also][]{Noterdaeme10co}. This implies that the molecular fractions $f$(\h2)
reported from the \h2 surveys should be considered as conservative lower limits for the \h2 components.

For two of the \h2-bearing DLAs with density measurements based on 
C~{\sc i} fine-structure excitation we derive an upper limit on the line of 
sight thickness of $\le 15$~pc. { This is consistent with the size estimate for the H$_2$-bearing gas in \zabs = 2.2377 DLA
towards Q1232+082 based on partial coverage \citep{Balashev11}.}

In principle, the presence of \h2 and 21-cm absorptions in a single component provides  
a unique combination to simultaneously constrain the variation
of the fine-structure constant ($\alpha$), the electron-to-proton mass ratio ($\mu$) and the proton G-factor. 
As shown here, DLAs with 21-cm 
and \h2 detections are rare. Even in these cases the presence of multiphase structure at parsec scale is evident, 
introducing velocity shifts
between the different absorption components that will affect the constraints on the variation of constants.

\section{acknowledgements}
We thank GBT, GMRT, VLBA and VLT staff for their support
during the observations and the anonymous referee for some
useful comments. We acknowledge the use of SDSS
spectra from the archive (http://www.sdss.org/).   
The National Radio Astronomy Observatory is a facility of the
National Science Foundation operated under cooperative agreement by
Associated Universities, Inc. 
VLBA data
were correlated using NRAO's implementation of the DiFX
software correlator that was developed as part of the Australian
Major National Research Facilities Programme and operated under
licence.
RS and PPJ gratefully acknowledge support from the Indo-French
Centre for the Promotion of Advanced Research (Centre Franco-Indien pour
la promotion de la recherche avanc\'ee) under Project N.4304-2.

\def\aj{AJ}%
\def\actaa{Acta Astron.}%
\def\araa{ARA\&A}%
\def\apj{ApJ}%
\def\apjl{ApJ}%
\def\apjs{ApJS}%
\def\ao{Appl.~Opt.}%
\def\apss{Ap\&SS}%
\def\aap{A\&A}%
\def\aapr{A\&A~Rev.}%
\def\aaps{A\&AS}%
\def\azh{AZh}%
\def\baas{BAAS}%
\def\bac{Bull. astr. Inst. Czechosl.}%
\def\caa{Chinese Astron. Astrophys.}%
\def\cjaa{Chinese J. Astron. Astrophys.}%
\def\icarus{Icarus}%
\def\jcap{J. Cosmology Astropart. Phys.}%
\def\jrasc{JRASC}%
\def\mnras{MNRAS}%
\def\memras{MmRAS}%
\def\na{New A}%
\def\nar{New A Rev.}%
\def\pasa{PASA}%
\def\pra{Phys.~Rev.~A}%
\def\prb{Phys.~Rev.~B}%
\def\prc{Phys.~Rev.~C}%
\def\prd{Phys.~Rev.~D}%
\def\pre{Phys.~Rev.~E}%
\def\prl{Phys.~Rev.~Lett.}%
\def\pasp{PASP}%
\def\pasj{PASJ}%
\def\qjras{QJRAS}%
\def\rmxaa{Rev. Mexicana Astron. Astrofis.}%
\def\skytel{S\&T}%
\def\solphys{Sol.~Phys.}%
\def\sovast{Soviet~Ast.}%
\def\ssr{Space~Sci.~Rev.}%
\def\zap{ZAp}%
\def\nat{Nature}%
\def\iaucirc{IAU~Circ.}%
\def\aplett{Astrophys.~Lett.}%
\def\apspr{Astrophys.~Space~Phys.~Res.}%
\def\bain{Bull.~Astron.~Inst.~Netherlands}%
\def\fcp{Fund.~Cosmic~Phys.}%
\def\gca{Geochim.~Cosmochim.~Acta}%
\def\grl{Geophys.~Res.~Lett.}%
\def\jcp{J.~Chem.~Phys.}%
\def\jgr{J.~Geophys.~Res.}%
\def\jqsrt{J.~Quant.~Spec.~Radiat.~Transf.}%
\def\memsai{Mem.~Soc.~Astron.~Italiana}%
\def\nphysa{Nucl.~Phys.~A}%
\def\physrep{Phys.~Rep.}%
\def\physscr{Phys.~Scr}%
\def\planss{Planet.~Space~Sci.}%
\def\procspie{Proc.~SPIE}%
\let\astap=\aap
\let\apjlett=\apjl
\let\apjsupp=\apjs
\let\applopt=\ao
\bibliographystyle{mn2e}
\bibliography{mybib}

\begin{thebibliography}{}

\bibitem[\protect\citeauthoryear{{Akerman}, {Ellison}, {Pettini} \&
  {Steidel}}{{Akerman} et~al.}{2005}]{Akerman05}
{Akerman} C.~J.,  {Ellison} S.~L.,  {Pettini} M.,    {Steidel} C.~C.,  2005,
  \aap, 440, 499

\bibitem[\protect\citeauthoryear{{Balashev}, {Petitjean}, {Ivanchik}, {Ledoux},
  {Srianand}, {Noterdaeme} \& {Varshalovich}}{{Balashev}
  et~al.}{2011}]{Balashev11}
{Balashev} S.~A.,  {Petitjean} P.,  {Ivanchik} A.~V.,  {Ledoux} C.,  {Srianand}
  R.,  {Noterdaeme} P.,    {Varshalovich} D.~A.,  2011, \mnras, pp 1400--+

\bibitem[\protect\citeauthoryear{{Briggs}, {Brinks} \& {Wolfe}}{{Briggs}
  et~al.}{1997}]{Briggs97}
{Briggs} F.~H.,  {Brinks} E.,    {Wolfe} A.~M.,  1997, \aj, 113, 467

\bibitem[\protect\citeauthoryear{{Briggs}, {Wolfe}, {Liszt}, {Davis} \&
  {Turner}}{{Briggs} et~al.}{1989}]{Briggs89}
{Briggs} F.~H.,  {Wolfe} A.~M.,  {Liszt} H.~S.,  {Davis} M.~M.,    {Turner}
  K.~L.,  1989, \apj, 341, 650

\bibitem[\protect\citeauthoryear{{Carilli}, {Lane}, {de Bruyn}, {Braun} \&
  {Miley}}{{Carilli} et~al.}{1996}]{Carilli96}
{Carilli} C.~L.,  {Lane} W.,  {de Bruyn} A.~G.,  {Braun} R.,    {Miley} G.~K.,
  1996, \aj, 111, 1830

\bibitem[\protect\citeauthoryear{{Curran}, {Tzanavaris}, {Darling}, {Whiting},
  {Webb}, {Bignell}, {Athreya} \& {Murphy}}{{Curran} et~al.}{2010}]{Curran10}
{Curran} S.~J.,  {Tzanavaris} P.,  {Darling} J.~K.,  {Whiting} M.~T.,  {Webb}
  J.~K.,  {Bignell} C.,  {Athreya} R.,    {Murphy} M.~T.,  2010, \mnras, 402,
  35

\bibitem[\protect\citeauthoryear{{Curran}, {Tzanavaris}, {Pihlstr{\"o}m} \&
  {Webb}}{{Curran} et~al.}{2007}]{Curran07}
{Curran} S.~J.,  {Tzanavaris} P.,  {Pihlstr{\"o}m} Y.~M.,    {Webb} J.~K.,
  2007, \mnras, 382, 1331

\bibitem[\protect\citeauthoryear{{Curran} \& {Webb}}{{Curran} \&
  {Webb}}{2006}]{Curran06}
{Curran} S.~J.,  {Webb} J.~K.,  2006, \mnras, 371, 356

\bibitem[\protect\citeauthoryear{{de Avillez} \& {Breitschwerdt}}{{de Avillez}
  \& {Breitschwerdt}}{2004}]{deavillez2004}
{de Avillez} M.~A.,  {Breitschwerdt} D.,  2004, \aap, 425, 899

\bibitem[\protect\citeauthoryear{{de Bruyn}, {O'Dea} \& {Baum}}{{de Bruyn}
  et~al.}{1996}]{DeBruyn96}
{de Bruyn} A.~G.,  {O'Dea} C.~P.,    {Baum} S.~A.,  1996, \aap, 305, 450

\bibitem[\protect\citeauthoryear{{Ellison}, {Pettini}, {Steidel} \&
  {Shapley}}{{Ellison} et~al.}{2001}]{Ellison01Q0201}
{Ellison} S.~L.,  {Pettini} M.,  {Steidel} C.~C.,    {Shapley} A.~E.,  2001,
  \apj, 549, 770

\bibitem[\protect\citeauthoryear{{Ellison}, {York}, {Pettini} \&
  {Kanekar}}{{Ellison} et~al.}{2008}]{Ellison08}
{Ellison} S.~L.,  {York} B.~A.,  {Pettini} M.,    {Kanekar} N.,  2008, \mnras,
  388, 1349

\bibitem[\protect\citeauthoryear{{Gupta}, {Salter}, {Saikia}, {Ghosh} \&
  {Jeyakumar}}{{Gupta} et~al.}{2006}]{Gupta06}
{Gupta} N.,  {Salter} C.~J.,  {Saikia} D.~J.,  {Ghosh} T.,    {Jeyakumar} S.,
  2006, \mnras, 373, 972

\bibitem[\protect\citeauthoryear{{Gupta}, {Srianand}, {Petitjean}, {Noterdaeme}
  \& {Saikia}}{{Gupta} et~al.}{2009}]{Gupta09}
{Gupta} N.,  {Srianand} R.,  {Petitjean} P.,  {Noterdaeme} P.,    {Saikia}
  D.~J.,  2009, \mnras, 398, 201

\bibitem[\protect\citeauthoryear{{Heinm{\"u}ller}, {Petitjean}, {Ledoux},
  {Caucci} \& {Srianand}}{{Heinm{\"u}ller} et~al.}{2006}]{Heinmuller06}
{Heinm{\"u}ller} J.,  {Petitjean} P.,  {Ledoux} C.,  {Caucci} S.,    {Srianand}
  R.,  2006, \aap, 449, 33

\bibitem[\protect\citeauthoryear{{Hirashita} \& {Ferrara}}{{Hirashita} \&
  {Ferrara}}{2005}]{Hirashita05}
{Hirashita} H.,  {Ferrara} A.,  2005, \mnras, 356, 1529

\bibitem[\protect\citeauthoryear{{Isobe}, {Feigelson} \& {Nelson}}{{Isobe}
  et~al.}{1986}]{Isobe86}
{Isobe} T.,  {Feigelson} E.~D.,    {Nelson} P.~I.,  1986, \apj, 306, 490

\bibitem[\protect\citeauthoryear{{Kanekar} \& {Chengalur}}{{Kanekar} \&
  {Chengalur}}{2003}]{Kanekar03}
{Kanekar} N.,  {Chengalur} J.~N.,  2003, \aap, 399, 857

\bibitem[\protect\citeauthoryear{{Kanekar}, {Chengalur} \& {Lane}}{{Kanekar}
  et~al.}{2007}]{Kanekar07}
{Kanekar} N.,  {Chengalur} J.~N.,    {Lane} W.~M.,  2007, \mnras, 375, 1528

\bibitem[\protect\citeauthoryear{{Kanekar}, {Lane}, {Momjian}, {Briggs} \&
  {Chengalur}}{{Kanekar} et~al.}{2009}]{Kanekar09vlba}
{Kanekar} N.,  {Lane} W.~M.,  {Momjian} E.,  {Briggs} F.~H.,    {Chengalur}
  J.~N.,  2009, \mnras, 394, L61

\bibitem[\protect\citeauthoryear{{Kanekar}, {Prochaska}, {Ellison} \&
  {Chengalur}}{{Kanekar} et~al.}{2009}]{Kanekar09mg2}
{Kanekar} N.,  {Prochaska} J.~X.,  {Ellison} S.~L.,    {Chengalur} J.~N.,
  2009, \mnras, 396, 385

\bibitem[\protect\citeauthoryear{{Kanekar}, {Smette}, {Briggs} \&
  {Chengalur}}{{Kanekar} et~al.}{2009}]{Kanekar09ts}
{Kanekar} N.,  {Smette} A.,  {Briggs} F.~H.,    {Chengalur} J.~N.,  2009,
  \apjl, 705, L40

\bibitem[\protect\citeauthoryear{{Kanekar}, {Subrahmanyan}, {Ellison}, {Lane}
  \& {Chengalur}}{{Kanekar} et~al.}{2006}]{Kanekar06}
{Kanekar} N.,  {Subrahmanyan} R.,  {Ellison} S.~L.,  {Lane} W.~M.,
  {Chengalur} J.~N.,  2006, \mnras, 370, L46

\bibitem[\protect\citeauthoryear{{Kulkarni} \& {Heiles}}{{Kulkarni} \&
  {Heiles}}{1988}]{Kulkarni88}
{Kulkarni} S.~R.,  {Heiles} C.,  1988, {Neutral hydrogen and the diffuse
  interstellar medium}.
pp 95--153

\bibitem[\protect\citeauthoryear{{Lane}, {Briggs} \& {Smette}}{{Lane}
  et~al.}{2000}]{Lane00}
{Lane} W.~M.,  {Briggs} F.~H.,    {Smette} A.,  2000, \apj, 532, 146

\bibitem[\protect\citeauthoryear{{Ledoux}, {Petitjean} \& {Srianand}}{{Ledoux}
  et~al.}{2003}]{Ledoux03}
{Ledoux} C.,  {Petitjean} P.,    {Srianand} R.,  2003, \mnras, 346, 209

\bibitem[\protect\citeauthoryear{{Levshakov} \& {Varshalovich}}{{Levshakov} \&
  {Varshalovich}}{1985}]{Levshakov85}
{Levshakov} S.~A.,  {Varshalovich} D.~A.,  1985, \mnras, 212, 517

\bibitem[\protect\citeauthoryear{{M{\o}ller}, {Fynbo} \& {Fall}}{{M{\o}ller}
  et~al.}{2004}]{Moller04}
{M{\o}ller} P.,  {Fynbo} J.~P.~U.,    {Fall} S.~M.,  2004, \aap, 422, L33

\bibitem[\protect\citeauthoryear{{Noterdaeme}, {Ledoux}, {Petitjean} \&
  {Srianand}}{{Noterdaeme} et~al.}{2008}]{Noterdaeme08}
{Noterdaeme} P.,  {Ledoux} C.,  {Petitjean} P.,    {Srianand} R.,  2008, \aap,
  481, 327

\bibitem[\protect\citeauthoryear{{Noterdaeme}, {Ledoux}, {Srianand},
  {Petitjean} \& {Lopez}}{{Noterdaeme} et~al.}{2009}]{Noterdaeme09co}
{Noterdaeme} P.,  {Ledoux} C.,  {Srianand} R.,  {Petitjean} P.,    {Lopez} S.,
  2009, \aap, 503, 765

\bibitem[\protect\citeauthoryear{{Noterdaeme}, {Petitjean}, {Ledoux},
  {L{\'o}pez}, {Srianand} \& {Vergani}}{{Noterdaeme}
  et~al.}{2010}]{Noterdaeme10co}
{Noterdaeme} P.,  {Petitjean} P.,  {Ledoux} C.,  {L{\'o}pez} S.,  {Srianand}
  R.,    {Vergani} S.~D.,  2010, \aap, 523, A80+

\bibitem[\protect\citeauthoryear{{Noterdaeme}, {Petitjean}, {Ledoux} \&
  {Srianand}}{{Noterdaeme} et~al.}{2009}]{Noterdaeme09dla}
{Noterdaeme} P.,  {Petitjean} P.,  {Ledoux} C.,    {Srianand} R.,  2009, \aap,
  505, 1087

\bibitem[\protect\citeauthoryear{{Noterdaeme}, {Petitjean}, {Srianand},
  {Ledoux} \& {L{\'o}pez}}{{Noterdaeme} et~al.}{2011}]{Noterdaeme11}
{Noterdaeme} P.,  {Petitjean} P.,  {Srianand} R.,  {Ledoux} C.,    {L{\'o}pez}
  S.,  2011, \aap, 526, L7+

\bibitem[\protect\citeauthoryear{{Petitjean}, {Bergeron} \&
  {Puget}}{{Petitjean} et~al.}{1992}]{Petitjean92}
{Petitjean} P.,  {Bergeron} J.,    {Puget} J.~L.,  1992, \aap, 265, 375

\bibitem[\protect\citeauthoryear{{Petitjean}, {Ledoux}, {Noterdaeme} \&
  {Srianand}}{{Petitjean} et~al.}{2006}]{Petitjean06}
{Petitjean} P.,  {Ledoux} C.,  {Noterdaeme} P.,    {Srianand} R.,  2006, \aap,
  456, L9

\bibitem[\protect\citeauthoryear{{Petitjean}, {Srianand} \&
  {Ledoux}}{{Petitjean} et~al.}{2000}]{Petitjean00}
{Petitjean} P.,  {Srianand} R.,    {Ledoux} C.,  2000, \aap, 364, L26

\bibitem[\protect\citeauthoryear{{Prochaska}, {Herbert-Fort} \&
  {Wolfe}}{{Prochaska} et~al.}{2005}]{Prochaska05}
{Prochaska} J.~X.,  {Herbert-Fort} S.,    {Wolfe} A.~M.,  2005, \apj, 635, 123

\bibitem[\protect\citeauthoryear{{Srianand}, {Gupta}, {Petitjean}, {Noterdaeme}
  \& {Ledoux}}{{Srianand} et~al.}{2010}]{Srianand10}
{Srianand} R.,  {Gupta} N.,  {Petitjean} P.,  {Noterdaeme} P.,    {Ledoux} C.,
  2010, \mnras, 405, 1888

\bibitem[\protect\citeauthoryear{{Srianand}, {Noterdaeme}, {Ledoux} \&
  {Petitjean}}{{Srianand} et~al.}{2008}]{Srianand08}
{Srianand} R.,  {Noterdaeme} P.,  {Ledoux} C.,    {Petitjean} P.,  2008, \aap,
  482, L39

\bibitem[\protect\citeauthoryear{{Srianand} \& {Petitjean}}{{Srianand} \&
  {Petitjean}}{1998}]{Srianand98}
{Srianand} R.,  {Petitjean} P.,  1998, \aap, 335, 33

\bibitem[\protect\citeauthoryear{{Srianand}, {Petitjean}, {Ledoux}, {Ferland}
  \& {Shaw}}{{Srianand} et~al.}{2005}]{Srianand05}
{Srianand} R.,  {Petitjean} P.,  {Ledoux} C.,  {Ferland} G.,    {Shaw} G.,
  2005, \mnras, 362, 549

\bibitem[\protect\citeauthoryear{{Wolfe}, {Briggs}, {Turnshek}, {Davis},
  {Smith} \& {Cohen}}{{Wolfe} et~al.}{1985}]{Wolfe85}
{Wolfe} A.~M.,  {Briggs} F.~H.,  {Turnshek} D.~A.,  {Davis} M.~M.,  {Smith}
  H.~E.,    {Cohen} R.~D.,  1985, \apjl, 294, L67

\bibitem[\protect\citeauthoryear{{Wolfe}, {Gawiser} \& {Prochaska}}{{Wolfe}
  et~al.}{2003}]{Wolfe03a}
{Wolfe} A.~M.,  {Gawiser} E.,    {Prochaska} J.~X.,  2003, \apj, 593, 235

\bibitem[\protect\citeauthoryear{{Wolfe}, {Gawiser} \& {Prochaska}}{{Wolfe}
  et~al.}{2005}]{wolfe05}
{Wolfe} A.~M.,  {Gawiser} E.,    {Prochaska} J.~X.,  2005, \araa, 43, 861

\bibitem[\protect\citeauthoryear{{Wolfe}, {Prochaska} \& {Gawiser}}{{Wolfe}
  et~al.}{2003}]{Wolfe03b}
{Wolfe} A.~M.,  {Prochaska} J.~X.,    {Gawiser} E.,  2003, \apj, 593, 215

\bibitem[\protect\citeauthoryear{{Wolfe}, {Prochaska}, {Jorgenson} \&
  {Rafelski}}{{Wolfe} et~al.}{2008}]{Wolfe08}
{Wolfe} A.~M.,  {Prochaska} J.~X.,  {Jorgenson} R.~A.,    {Rafelski} M.,  2008,
  \apj, 681, 881

\bibitem[\protect\citeauthoryear{{Wolfire}, {Hollenbach}, {McKee}, {Tielens} \&
  {Bakes}}{{Wolfire} et~al.}{1995}]{Wolfire95}
{Wolfire} M.~G.,  {Hollenbach} D.,  {McKee} C.~F.,  {Tielens} A.~G.~G.~M.,
  {Bakes} E.~L.~O.,  1995, \apj, 443, 152

\bibitem[\protect\citeauthoryear{{York}, {Kanekar}, {Ellison} \&
  {Pettini}}{{York} et~al.}{2007}]{York07}
{York} B.~A.,  {Kanekar} N.,  {Ellison} S.~L.,    {Pettini} M.,  2007, \mnras,
  382, L53

\bibitem[\protect\citeauthoryear{{Zwaan} \& {Prochaska}}{{Zwaan} \&
  {Prochaska}}{2006}]{Zwaan06}
{Zwaan} M.~A.,  {Prochaska} J.~X.,  2006, \apj, 643, 675

\end{thebibliography}

\end{document}